\documentclass[aps,prl,reprint,superscriptaddress,nobibnotes]{revtex4-2}
\usepackage{mathrsfs}
\usepackage{amsmath,gensymb}
\usepackage{amsfonts}
\usepackage{amssymb}
\usepackage{amsthm}
\usepackage{graphicx}
\usepackage{natbib}
\usepackage{xcolor}
\usepackage{hyperref}
\usepackage{bm}
\usepackage[caption=false]{subfig}
\usepackage{verbatim}
\usepackage{siunitx}

\begin{document}
\title{Magneto-Chiral Anisotropy in Josephson Diode Effect of All-Metallic Lateral Junctions with Interfacial Rashba Spin-Orbit Coupling}

\author{Maximilian Mangold}
\affiliation{School of Natural Sciences, Technical University of Munich, 85748 Garching b. Munich, Germany}
\affiliation{Center for Quantum Engineering (ZQE), Technical University of Munich, 85748 Garching b. Munich, Germany}
\author{Lorenz Bauriedl}
\affiliation{Department of Physics, University of Regensburg, 93040 Regensburg, Germany}
\affiliation{Halle-Berlin-Regensburg Cluster of Excellence CCE, University of Regensburg, 93040 Regensburg, Germany}
\author{Johanna Berger}
\affiliation{Department of Physics, University of Regensburg, 93040 Regensburg, Germany}
\affiliation{Halle-Berlin-Regensburg Cluster of Excellence CCE, University of Regensburg, 93040 Regensburg, Germany}
\author{Chang Yu-Cheng}
\affiliation{Pico Group, QTF Centre of Excellence, Department of Applied Physics, Aalto University, P.O. Box 15100, FI-00076 Aalto, Finland}
\author{Thomas N.G. Meier}
\affiliation{School of Natural Sciences, Technical University of Munich, 85748 Garching b. Munich, Germany}
\affiliation{Center for Quantum Engineering (ZQE), Technical University of Munich, 85748 Garching b. Munich, Germany}
\author{Matthias Kronseder}
\affiliation{Department of Physics, University of Regensburg, 93040 Regensburg, Germany}
\affiliation{Halle-Berlin-Regensburg Cluster of Excellence CCE, University of Regensburg, 93040 Regensburg, Germany}
\author{Pertti Hakonen}
\affiliation{Low Temperature Laboratory, Department of Applied Physics, Aalto University, P.O. Box 15100, FI-00076 Espoo, Finland}
\author{Christian H. Back}
\affiliation{School of Natural Sciences, Technical University of Munich, 85748 Garching b. Munich, Germany}
\affiliation{Center for Quantum Engineering (ZQE), Technical University of Munich, 85748 Garching b. Munich, Germany}
\author{Christoph Strunk}
\affiliation{Department of Physics, University of Regensburg, 93040 Regensburg, Germany}
\affiliation{Halle-Berlin-Regensburg Cluster of Excellence CCE, University of Regensburg, 93040 Regensburg, Germany}
\author{Dhavala Suri}
\email[Corresponding Author: ]{dsuri@iisc.ac.in}
\affiliation{School of Natural Sciences, Technical University of Munich, 85748 Garching b. Munich, Germany}
\affiliation{Centre for Nanoscience and Engineering, Indian Institute of Science, Bengaluru 560012, India}

\begin{abstract}
We explore the role of interfacial Rashba spin-orbit coupling (SOC) for the Josephson diode effect in all-metal diffusive Josephson junctions. Devices with Fe/Pt and Cu/Pt weak links between Nb leads reveal a Josephson diode effect in an in-plane magnetic field with magneto-chiral anisotropy according to the point symmetry of Rashba SOC. The Rashba SOC originates from inversion symmetry breaking at the metal-metal interfaces. A control sample with a plain Cu-layer as weak link, in contrast, exhibits an axis-symmetric diode effect.
The Fraunhofer patterns display an apparent inverted hysteresis which can be traced back to stray fields resulting from the conventional hysteretic vortex pinning in the Nb contacts.

\end{abstract}
	
\maketitle

Hybrid Josephson junctions (JJs) have enabled major advances in fundamental research of superconductivity and are central to contemporary developments in quantum technology.
Milestone experiments have lead to novel qubit architectures~\cite{Aguado_2017, Aguado_2020, DasSarma_2023}, memory devices based on magnetic JJs~\cite{BirgeSatchell_2024}, and spin-polarized supercurrents~\cite{Eschrig_2015, LinderRobinson_2015}.

The recent discovery of the Josephson diode effect (JDE)~\cite{Baumgartner_2022, Pal_2022, Wu_2022, Costa_2023, Amundsen_2024} has added a new circuit element to future superconducting electronics~\cite{Golod_2022, Suri_2022, Hou_2023, Ingla_2025, Castellani_2025}. The JDE can occur when time-reversal symmetry (TRS) and inversion symmetries are broken and is reflected in an asymmetry of positive and negative critical current. Such asymmetry is a manifestation of magneto-chiral anisotropy, introduced by Ref.~\cite{Rikken_2005}. So far, the diode effect in JJs was mostly observed in semiconductor weak links like InAs or InSb with strong Rashba spin-orbit coupling (SOC)~\cite{Baumgartner_2022, Turini_2022}.

\begin{figure*}[tbh]
\centering
\includegraphics[width=7in]{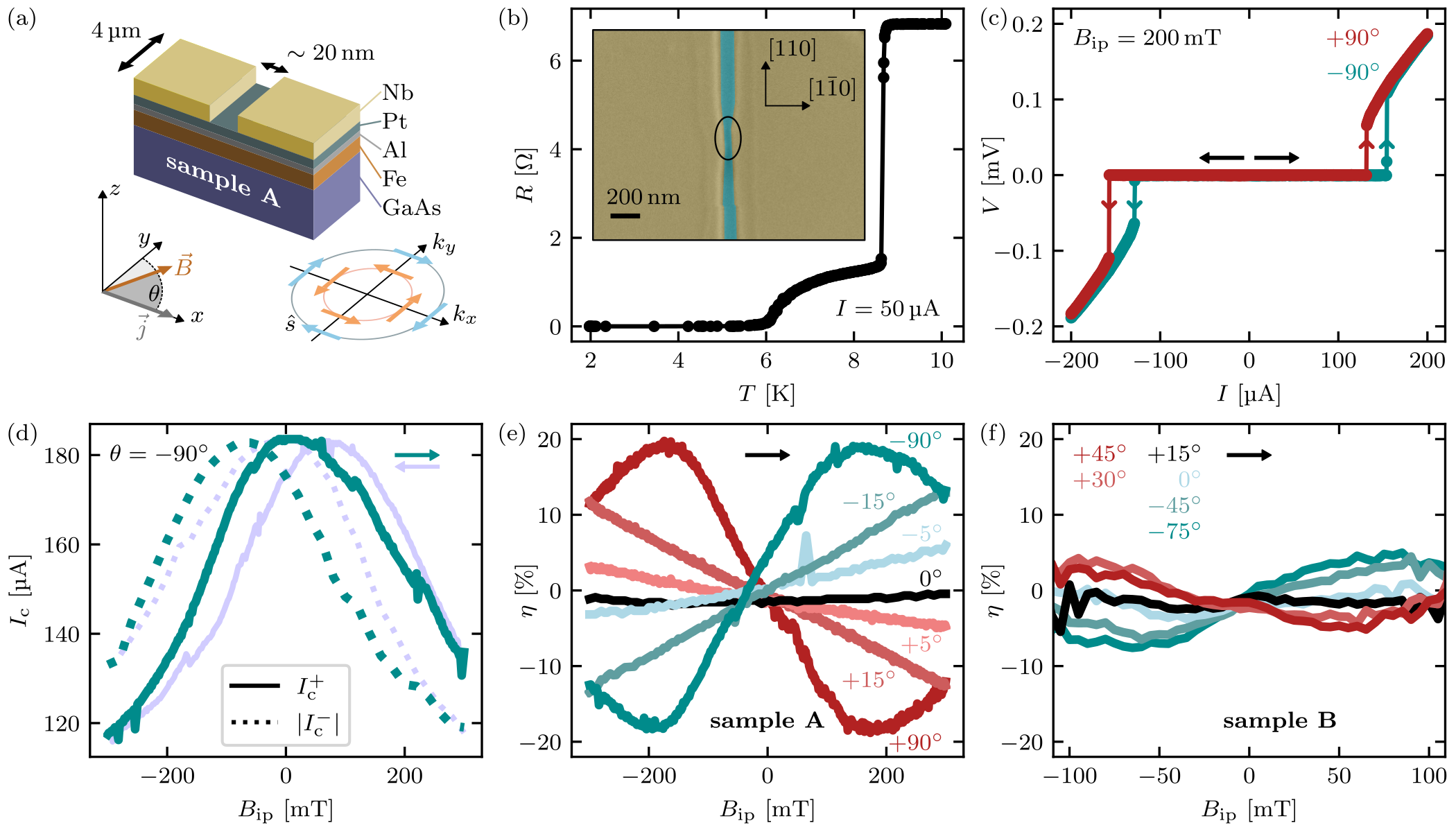}
\caption{Diode efficiency under $B_\textrm{ip}$. (a) Device and layer sequence of sample~A. The direction of positive applied current defines the $x$-axis. The direction of $B_\textrm{ip}$ applied in panels (c-f) is indicated by the coordinate system. $\theta$ measures the angle between $B_\textrm{ip}$ and the direction of the positive current density $\vec{j}$. The bottom right sketch depicts the chiral spin texture of the Rashba SOC at the Fermi level in momentum space. (b) Temperature-dependence of the resistance of sample~A measured at $I = \SI{50}{\micro A}$ in zero magnetic field. Inset: SEM false-color image of the device. The Nb leads are highlighted in yellow, the blue area represents the junction. At the narrowest point, encircled in the SEM image, the gap measures around \SI{20}{nm} in length and $\sim\SI{180}{nm}$ in width. The indicated crystallographic axes refer to the GaAs substrate. (c) IV curves of sample~A taken at $B_\textrm{ip} = \SI{200}{mT}$ applied at $\theta = +90^\circ$ (red) and $-90^\circ$ (blue) in forward field sweep direction. Arrows indicate the sweep direction of $I$, ramping from zero to \SI{+200}{\micro A} and from zero to \SI{-200}{\micro A}. (d) In-plane field-dependence of $I_\textrm{c}$ of sample~A for fixed $\theta = -90^\circ$. Absolute values of $I_\textrm{c}$ at positive ($I_\textrm{c}^+$) and negative ($I_\textrm{c}^-$) bias are indicated by solid and dashed lines, respectively. The sweep direction of the applied field is indicated by the colored arrows. (e) Diode efficiency $\eta$ of sample~A as a function of $B_\textrm{ip}$ at various in-plane angles $\theta$. The field is swept from negative to positive values. The $I_\textrm{c}$ curves producing the depicted $\eta(B_\textrm{ip},\theta)$ are shown in the Supplemental Material~\cite{SupplMat}. Measurements shown in panels (c-e) are taken at \SI{30}{mK}. (f) $\eta(B_\textrm{ip})$ of sample~B at various angles $\theta$, measured at \SI{1.8}{K}. The curves at $\theta = +30^\circ$ and $+15^\circ$ are shifted to create a crossing at zero applied field, see Supplemental Material~\cite{SupplMat}. All curves in panels (e-f) are acquired in forward field sweep direction.}
\label{intro}
\end{figure*}

In bi-layer metallic films, inversion symmetry is broken by the charge transfer between the metal films, leading to a short-ranged, but strong interface electric field and thus to a Rashba-type Hamiltonian $\mathcal{H}_\textrm{R} = \alpha(\vec{k}\times\hat{z})\cdot \hat{s}$, where $\vec{k}$ and $\hat{s}$ are the electron's momentum and spin, respectively, and $\alpha$ is the Rashba constant~\cite{Rashba_1960, Bychkov_1984, Manchon_2015, Ganichev_2014}. In momentum space, a characteristic circular spin texture emerges at the Fermi level~\cite{GambardellaMiron_2011} [Fig.~\ref{intro}(a)]. Interfacial Rashba SOC at metal-metal interfaces is reflected, {\it e.g.}, in spin-orbit torques~\cite{Manchon_2008, Miron_2010}, and was demonstrated at the Cu/Pt interface by spin-injection techniques~\cite{Sanchez_2014, Yu_2020}. Strong interfacial spin-orbit interactions have also been observed in Fe/Pt on GaAs in spin-pumping measurements driven by ferromagnetic resonance~\cite{Chen_2016, Chen_2024}.

In contrast, studies on Josephson junctions of metallic weak links with interfacial Rashba SOC have been sparse. Senapati~{\it et al.}~\cite{Senapati_2023} investigated Cu/Pt bi-layers as weak link in Nb-based JJs. They reported an apparent inverted hysteresis where the central Fraunhofer lobe is located at negative fields for forward field sweeps and vice versa. Similar inverted hysteresis was also found in Al-based JJs with Fe-doped InAs as weak links~\cite{Nakamura_2019} and in epitaxial CoSi$_2$/TiSi$_2$~\cite{Chiu_2021} normal metal/superconductor junctions. To date, no consensus on the origin of the inverted hysteresis has been reached.

In this Letter, we explore the experimental manifestation of the interfacial Rashba SOC in the JDE, demonstrating the possibility to use the JDE as a probe for interfacial Rashba SOC. We investigate the effect of interfacial SOC on the supercurrent in Nb-based lateral JJs with metallic weak links. The JDE is observed only for bi-layer Fe/Pt and Cu/Pt. There, the breaking of inversion symmetry leads to interfacial Rashba SOC with the typical two-fold symmetry with respect to the orientation of the in-plane magnetic field. For Cu weak links, in contrast, the diode efficiency has an even symmetry.
All devices exhibit an apparent inverted hysteresis in out-of-plane fields that can be understood as a result of hysteretic vortex pinning in the Nb terminals.

We first examine an epitaxial heterostructure of GaAs(001)/Fe(\SI{2.9}{nm})/\,Al(\SI{1}{nm})/\,Pt(\SI{5}{nm})/\,Nb(\SI{50}{nm}), (sample~A). The specific stack is chosen for its strong interfacial SOC arising at both the GaAs/Fe and the Fe/Al/Pt interface~\cite{Chen_2016, Chen_2024}. The thin layer of Al serves as a spacer, preventing magnetic proximity polarization of the Pt layer while remaining transparent for charge and spin transport~\cite{Chen_2024}. A lateral Nb JJ is realized atop the Pt film by etching a narrow gap into the Nb layer. Fig.~\ref{intro}(a) depicts the stack schematically, and an SEM image is shown in the inset of Fig.~\ref{intro}(b).

We study the device resistance $R$ as a function of temperature and observe a two-step behavior characteristic of a hybrid JJ. At $T\approx \SI{8.6}{K}$, the Nb electrodes become superconducting first, followed by a broad transition to $R=0$ near $T\approx \SI{6.0}{K}$ [see Fig.~\ref{intro}(b)]. To the best of our knowledge, this is the first instance of an experimental realization of a lateral Josephson junction with a Fe weak link~\cite{Robinson_2006, Piano_2007}. To investigate the signatures of Rashba-type SOC in the present device, we study the critical current $I_\textrm{c}$ under a magnetic field $B_\textrm{ip}$ applied in the sample plane. To avoid heating-induced hysteresis, IV curves are acquired by ramping the applied current $I$ first from zero to \SI{+200}{\micro A} and then from zero to \SI{-200}{\micro A} while measuring the voltage drop $V$ across the junction in a 4-point geometry. At \SI{30}{mK}, $I_\textrm{c}R_\textrm{n} \approx \SI{3.0}{mV}$ which is comparable to the superconducting gap $\Delta$ and the Thouless energy $E_\textrm{Th} \approx \SI{1.6}{meV} \approx 1.2 \Delta$ (see Supplemental Material~\cite{SupplMat}). This is consistent with both the short and long junction limits~\cite{Dubos_2001}.
Fig.~\ref{intro}(c) displays two IV curves taken at $B_\textrm{ip} = \SI{200}{mT}$ applied at an angle $\theta = +90^\circ$ (red) and $-90^\circ$ (blue) with respect to $I$. 
The difference in $I_\textrm{c}$ under positive ($I_\textrm{c}^+$) and negative bias ($I_\textrm{c}^-$) is quantified through the diode efficiency~\cite{He_2022}
$\eta = 2({I_\textrm{c}^+ - |I_\textrm{c}^-|})/({I_\textrm{c}^+ + |I_\textrm{c}^-|)} \times 100\%$.
Here $\eta = \SI{-17.8}{\%}$ and \SI{18.0}{\%} at $\theta = \pm 90^\circ$, respectively. Fig.~\ref{intro}(d) depicts the magnetic field dependence of $I_\textrm{c}^+$ and $|I_\textrm{c}^-|$ measured at $\theta = -90^\circ$ in both forward (from $-B_{\textrm{ip}}$ to $+B_{\textrm{ip}}$, blue) and backward (from $+B_{\textrm{ip}}$ to $-B_{\textrm{ip}}$, lavender) field sweep direction. All curves show the expected bell-like shape of a JJ under $B_\textrm{ip}$\cite{Tinkham_2004, Suominen_2017, Turini_2022}. A considerable JDE is observed between $I_\textrm{c}^+$ and $|I_\textrm{c}^-|$ for both sweep directions. The resulting diode efficiency $\eta(B_{\textrm{ip}})$ is depicted in Fig.~\ref{intro}(e) for various angles $\theta$ in forward field-sweep direction. At $\theta = -90^\circ$ (\textit{i.e.} $\vec{B}~\perp~\vec{j}$), it features the characteristic shape of a diode effect~\cite{Suri_2022, Bauriedl_2022, Turini_2022}. A maximum diode efficiency of $\SI{\pm 20}{\%}$ is reached around $\SI{\pm 175}{mT}$, and $\eta(B_\textrm{ip})$ exhibits positive polarity (positive cusp at positive field and vice versa). As the field direction is rotated towards $+90^\circ$, the diode efficiency is gradually reduced, vanishes and changes sign at $\theta = 0^\circ$ (\textit{i.e.} $\vec{B}~||~\vec{j}$) before returning to its initial strength with opposite polarity at $+90^\circ$, constituting the typical $2\pi$-periodic magneto-chiral anisotropy. Since the reversal of sign in $\eta$ occurs at $\theta = 0^\circ$, the polarity matches that of an interfacial Rashba SOC inside the weak link.

We repeat the same measurements on non-magnetic weak links in order to substantiate the influence of Rashba SOC and rule out the in-plane $\langle 110 \rangle$ uniaxial magnetic anisotropy of GaAs/Fe~\cite{Bayreuther_2012} as the source of the diode effect. Sputtered Si/SiO$_2$/Cu(\SI{50}{nm})/Pt(\SI{3.5}{nm})/Nb(\SI{40}{nm}) films (sample~B) induce Rashba SOC at the Cu/Pt interface~\cite{Sanchez_2014, Yu_2020}. A lateral JJ is fabricated following the same methods as for the first sample. The device's critical temperature is near \SI{5}{K}, and IV characteristics for $T<\SI{5}{K}$ yield $I_\textrm{c}R_\textrm{n} \approx \SI{26}{\micro V}$ (see Supplemental Material~\cite{SupplMat}), which is comparable to literature values for similar Cu weak links~\cite{Senapati_2023, Garcia_2009}.
Similarly to sample~A, we record IV curves under $B_\textrm{ip}$ at various angles $\theta$. Before each field sweep the sample temperature is brought to \SI{9}{K} -- which is above $T_\textrm{c}$ of the Nb leads -- and field-cooled at $\pm \SI{200}{mT}$. This procedure eliminates the hysteretic effect of previous field sweeps as described below, and ensures a well-defined initial state. For sample~A, this step was not necessary due to the larger diode efficiency. We extract $I_\textrm{c}$ and calculate the diode efficiency $\eta$ shown in Fig.~\ref{intro}(f). Compared to Fig.~\ref{intro}(e), the maximum diode efficiency is smaller. Nevertheless, the characteristic Rashba-like magnetochiral anisotropy is observed, where the polarity of $\eta(B_\textrm{ip})$ gradually changes from positive at $\theta = -75^\circ$ to negative at $+45^\circ$, with the sign change occurring around $0^\circ$. A larger range of angles was not accessible experimentally. Note that the traces at $\theta = +30^\circ$ and $+15^\circ$ in Fig.~\ref{intro}(f) are slightly shifted to cross with the other curves at zero magnetic field (see Supplemental Material~\cite{SupplMat} for details). Importantly, the polarity of the diode effect is the same for both Fe/Pt and Cu/Pt systems. This is consistent as Fe and Cu share a similar work function ($W_\textrm{Fe}\approx\SI{4.7}{eV}$, $W_\textrm{Cu}\sim\SI{4.5}{eV}$), thus producing the same sign for the Rashba SOC when interfaced with Pt ($W_\textrm{Pt}\sim\SI{5.6}{eV}$)~\cite{Haynes_2016}. The similarity of the results for both samples suggests that the effect is most likely related to Rashba-type SOC rather than the ferromagnetic Fe film of sample~A.

Next, we analyze the diode efficiency in JJs where weak links are \SI{50}{nm} Cu films with no metal-metal interfaces and negligible bulk SOC (sample~C)~\cite{Du_2014, Sinova_2015}. Details on the fabrication process and device characterization are given in the Supplemental Material~\cite{SupplMat}. Also in this device, a diode efficiency between $-\SI{50}{\%}$ and $+\SI{20}{\%}$ is observed. However, in contrast to the bi-layer devices with Rashba interfaces where $\eta(\theta)$ displays a point symmetric shape, the diode efficiency of sample~C is approximately even in $\theta$ (see Supplemental Material~\cite{SupplMat}). In this case, the Rashba SOC can be excluded as the origin of the diode effect. The assignment to an alternative mechanism requires further investigation.

A central result of this article is that only for devices with metal-metal interfaces (samples~A and~B), the observed angle dependence points towards Rashba SOC. For sample~A (GaAs/Fe/Al/Pt), where strong Rashba SOC has been reported~\cite{Chen_2016, Chen_2024}, the diode effect is more pronounced than for sample~B (Cu/Pt) which is known to have weaker SOC~\cite{Sanchez_2014, Yu_2020}. This strongly supports the conclusion that the JDE is tuned by purely changing the interfacial configuration, while bulk SOC does not play a significant role. This is consistent with observations by Senapati \emph{et al.}~\cite{Senapati_2023} on comparable structures.

In addition to the JDE, a substantial hysteretic shift in magnetic field is observed in Fig.~\ref{intro}(d) when comparing the curves of $I_\textrm{c}^+$ measured with opposite field sweep directions (solid lines). The dashed traces representing $|I_\textrm{c}^-|$ are shifted in the same fashion. Remarkably, this shift does not conform with conventional magnetic hysteresis; instead it is characterized by an inverted order. For all samples, we also observe a notable inversely hysteretic shift in a perpendicular field $B_z$ between the forward and backward sweep direction, as analyzed in detail in the End Matter. We propose a mechanism based on the hysteretic vortex pinning inside the Nb leads~\cite{Bean_1962}.

\begin{figure}[t!]
\centering
\includegraphics[width=3.4in]{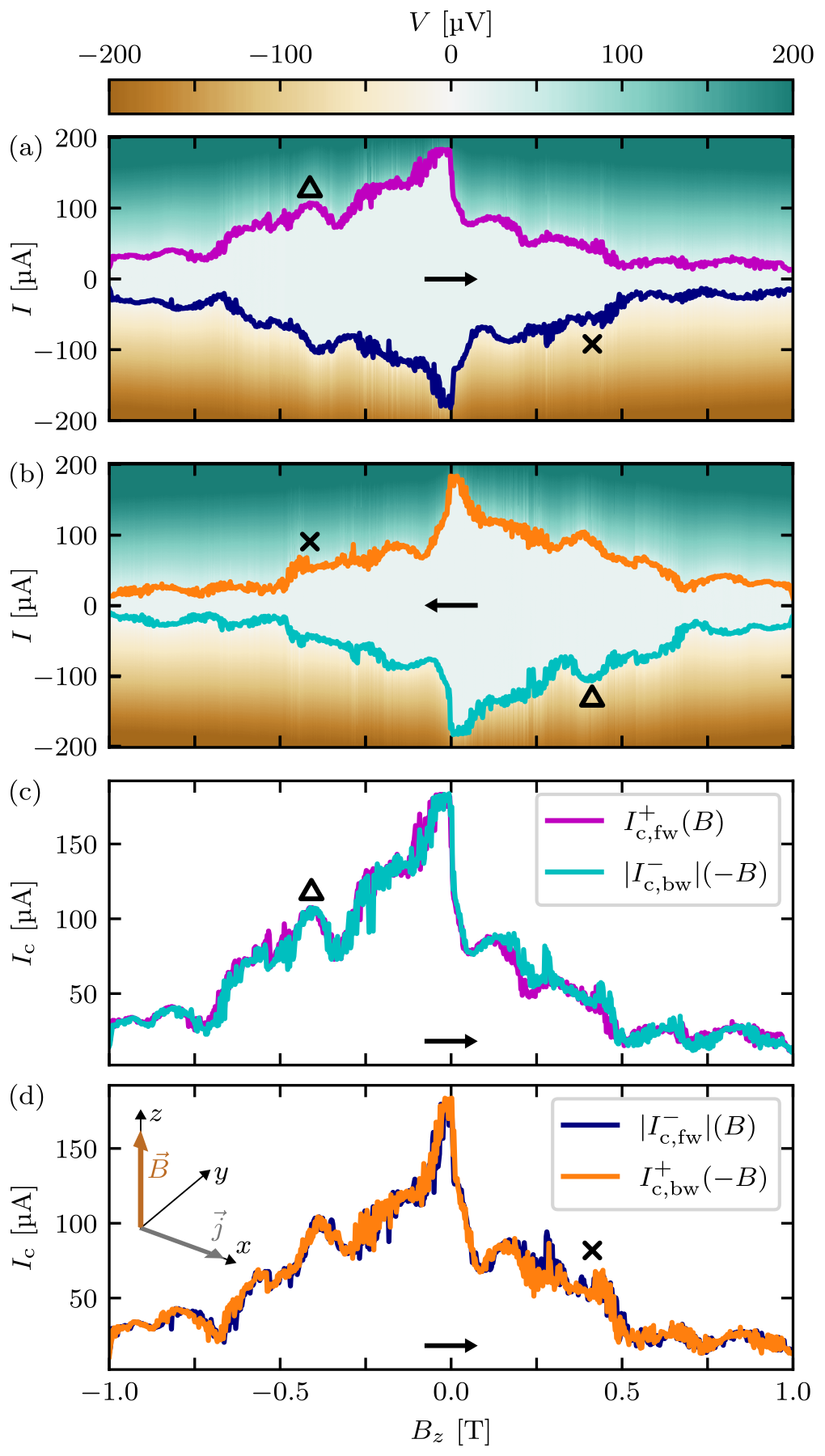}
\caption{Interference pattern of sample~A under $B_z$ in (a) forward and (b) backward field sweep direction, as indicated by the arrows. The extracted $I_\textrm{c}$ curves are overlaid in different colors. The triangles indicate a feature that is reproduced only by the inversion of current, field and field sweep direction. The cross marks the position that is obtained by inverting only field and current direction. (c) Comparison of $I_\textrm{c,fw}^+$ [see top half of data in panel (a)] with $|I_\textrm{c,bw}^-|$ [bottom half of panel (b)]. Note that the field axis of $|I_\textrm{c,bw}^-|$ is flipped to match the forward field sweep direction. (d) Equivalent comparison of $|I_\textrm{c,fw}^-|$ [bottom half of panel (a)] and $I_\textrm{c,bw}^+$ [top half of panel (b)]. The measurements are taken at \SI{30}{mK}. The inset of panel (d) shows the direction of the applied field.}
\label{oop}
\end{figure}

Finally, we investigate the supercurrent interference of sample~A in a perpendicular magnetic field. Figures.~\ref{oop}(a) and (b) display the interference pattern of the junction's IV characteristic as a function of $B_z$ in forward and backward field sweep direction, respectively. Notably, $I_\textrm{c}(B_z)$ exhibits a pronounced asymmetry about $B_z = 0$. The deviation from the standard Fraunhofer pattern can be attributed to the spatial inhomogeneity of the junction's critical current density owing to fabrication imperfections. The apparent noise is related to spontaneous flux jumps in the Nb leads (see End Matter for more details).
Upon comparing the forward and backward sweeps, a similar inverted hysteresis is present, see Supplemental Material~\cite{SupplMat}. 

Interestingly, the combined transformations $I_\textrm{c}^+ \rightarrow I_\textrm{c}^-$ and $+B_z \rightarrow -B_z$ do not lead to a match of the corresponding curves that is expected from TRS. The reason is that the hysteretic stray fields from the pinned vortices must be included to reverse {\it all} magnetic fields. Thus, a proper check for TRS also has to reverse the field sweep direction, not only $I_\textrm{c}$ and the applied field $B_z$.
As an example, the value of $I_\textrm{c,fw}^+$ near $-\SI{0.4}{T}$ in Fig.~\ref{oop}(a) marked by the triangle is not reproduced in $-I_\textrm{c,fw}^-$ at $+\SI{0.4}{T}$, as indicated by the cross. However, it is equal to $-I_\textrm{c,bw}^-$ at $+\SI{0.4}{T}$ {\it and} the opposite sweep direction, as highlighted by the triangle in Fig.~\ref{oop}(b). For a direct comparison, $I_\textrm{c,fw}^+$ and $|I_\textrm{c,bw}^-|$ are plotted together in Fig.~\ref{oop}(c). For the latter, the field axis is reversed to account for the backward sweep direction. The curves overlap in all prominent features. Similarly, $|I_\textrm{c,fw}^-|$ and $I_\textrm{c,bw}^+$ can be matched, as shown in Fig.~\ref{oop}(d). In the Supplemental Material~\cite{SupplMat} it is demonstrated that other combinations do not result in an adequate overlap.

To summarize, Nb-based Josephson junctions comprising Fe/Al/Pt and Cu/Pt weak links reveal the signatures of Rashba SOC in diffusive, all-metallic Josephson junctions. An apparently inverted hysteresis is traced back to the standard hysteresis of pinned vortices in the Nb leads. The consideration of this pinning effect is crucial in superconducting electronics where the phase bias is a key parameter. \\

\begin{acknowledgments}
The authors thank Lin Chen, Nicola Paradiso and Jukka Pekola for insightful discussions. This work was funded by Deutsche Forschungsgemeinschaft (DFG, German Research Foundation) through Project-ID 314695032 - SFB 1277 (Subprojects A08 and B08), and by the European Union’s Horizon 2020 Research and Innovation Programme, under Grant Agreement No. 824109 (EMP). DS acknowledges funding from Marie Sklodowska-Curie postdoctoral fellowship from EuroTechPostdoc2. DS thanks INOXCVA, INOX Airproducts and Infosys Foundation for financial support. The experimental work benefited from the Aalto University OtaNano/LTL infrastructure.  \\
\end{acknowledgments}

\textit{Author contribution statement} --  CHB, CS and DS conceptualized the experiments. MM and DS made the devices. MM, DS, LB, JB and CYC performed the measurements. MK and TNGM provided samples. MM, CS and DS analyzed the data. MM, CS and DS wrote the manuscript with inputs from CHB, MK, and PH. All authors contributed to the discussions.   \\

\textit{Data availability} -- The data that support the findings of this article are openly available~\cite{Zenodo}.

\bibliography{00_arXiv_references.bib}

\appendix
\section{End Matter}

Here we discuss the origin of the peculiar phenomenon of inverted hysteresis, which is explained below in terms of a consequence of the stray field of conventionally pinned vortices in hard superconductors. Previous reports on the inverted hysteresis of critical currents in JJs of type-II superconductors have been linked to the presence of Rashba SOC inside the JJ~\cite{Senapati_2023} as well as triplet-pair superconductivity~\cite{Nakamura_2019, Chiu_2021}. Since we find inverted hysteresis in all junctions with and without SOC, the effect appears not to be related to SOC. A more conventional explanation takes into account hysteretic vortex pinning as described by Bean's critical state model~\cite{Bean_1962}. Depending on the sequence of the field sweep procedure, the pinned vortices can create a stray field that opposes the externally applied field $B_z$ and produces a net zero flux through the JJ before $B_z$ is swept past zero.

\begin{figure*}[tbh]
\centering
\includegraphics[width=7in]{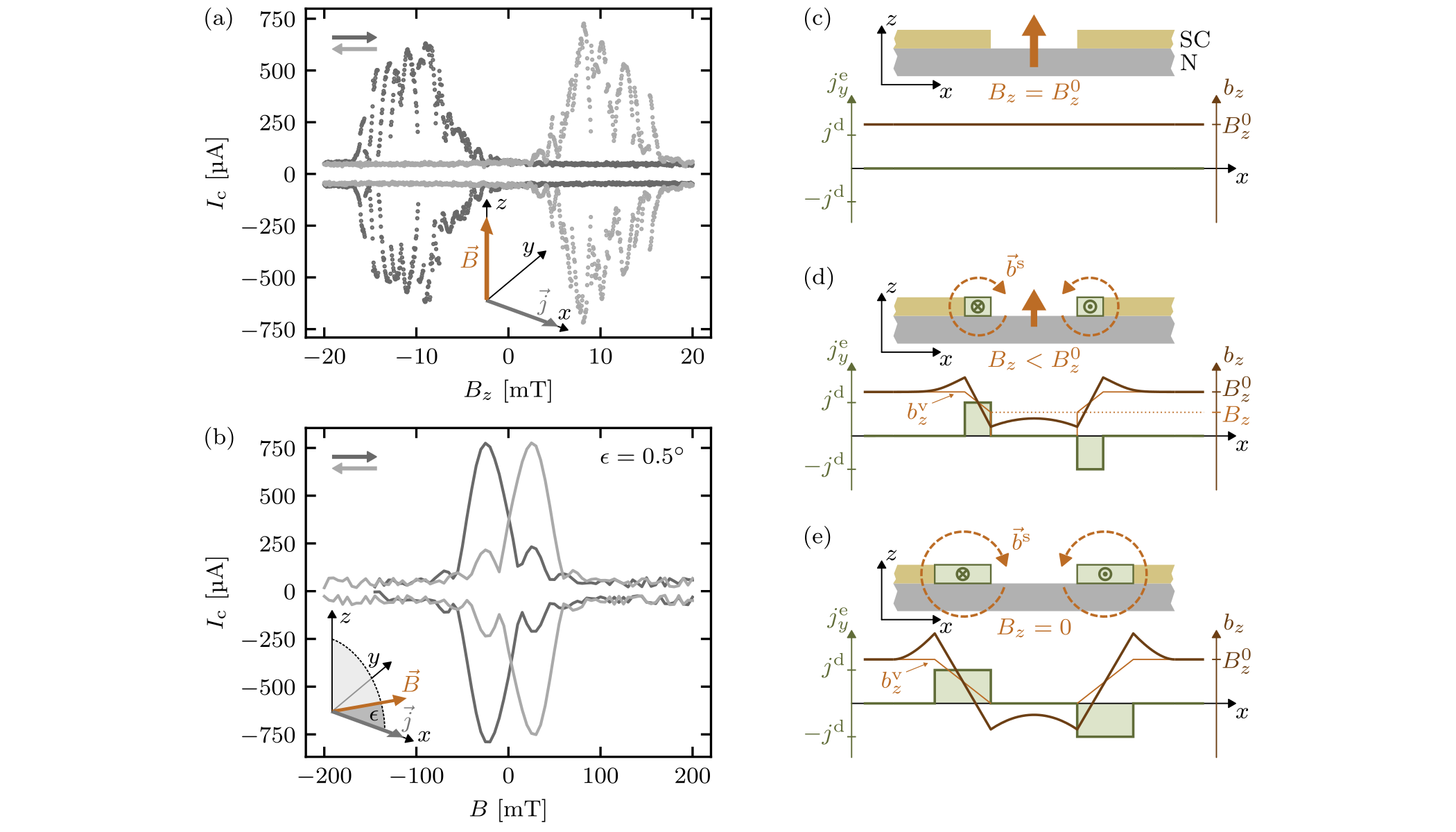}
\caption{Inverted hysteresis in sample~B. (a) $I_\textrm{c}(B_z)$ in forward (dark) and backward (light) field sweep direction. (b) $I_\textrm{c}(B)$ for $B$ applied along the current direction with a small misalignment of $\epsilon=0.5^\circ$ from the sample plane. All measurements are taken at \SI{1.8}{K}. (c-e) Schematic cross-section of a generic JJ in a superconductor (SC) on top of a normal conductor (N). $b_z(x)$ denotes the local magnetic field given by the external field $B_z$, the field $b^\textrm{v}_z$ generated by the vortex distribution inside the Nb leads, and the stray magnetic field $\vec{b}^\textrm{s}$ induced by the edge currents $j^\textrm{e}_y$. The corresponding profiles of $b_z(x)$, $b^\textrm{v}_z(x)$ and $j^\textrm{e}_y(x)$ according to the critical state model~\cite{Bean_1962} are shown at different stages during a downward field sweep after field-cooling at $B_z^0$. For better visibility, the contribution of $b^\textrm{s}_z$ is exaggerated. For the discussion of the field profile, see Supplemental Material~\cite{SupplMat}.}
\label{InvHyst}
\end{figure*}

Fig.~\ref{InvHyst}(a) shows $I_\textrm{c}(B_z)$ for sample~B where $B_z$ is swept from \SI{-20}{mT} to \SI{+20}{mT} (forward, dark gray) and from \SI{+20}{mT} to \SI{-20}{mT} (backward, light gray) after field-cooling at $\pm \SI{20}{mT}$. An inversely hysteretic shift of the interference patterns in forward and backward field sweep direction of around $\mp\SI{10}{mT}$ is observed. Experimental artifacts from remnant fields stored in the coils are ruled out as the origin of the inverted hysteresis in the Supplemental Material~\cite{SupplMat}. The $I_\textrm{c}(B_z)$ traces feature a broad inversely hysteretic bump with several jumps of $I_\textrm{c}$ while the field is swept. The switching pattern varies for several repetitions of the sweep (see Supplemental Material~\cite{SupplMat}). We assign the discontinuities to flux jumps in the junctions resulting from spontaneous changes in the stray magnetic field of vortices pinned in the Nb strips near the junction. The instabilities in the stray field are caused by random depinning events as the applied field changes. Between these jumps, Fig.~\ref{InvHyst}(a) reveals a smooth change of $I_\textrm{c}$ over 1-3~mT. In this way, short intervals of an intrinsically clean Fraunhofer pattern are probed multiple times, until the flux in the junction exceeds a few flux quanta and $I_\textrm{c}(B_z)$ is featureless (see Supplemental Material~\cite{SupplMat} for more details). Flux jumps in $I_\textrm{c}(B_z)$ are also seen for sample~A in Fig.~\ref{oop} on the few \SI{}{mT}-scale across the whole field range, as demonstrated in the Supplemental Material~\cite{SupplMat}.

Flux jumps are avoided when applying the magnetic field essentially parallel, but with a small misalignment angle $\epsilon=0.5^\circ$ from the sample surface. In this configuration, vortices thread the Nb film over the full width of the contacts. Hence, each vortex is expected to contribute only a small amount $\phi = \phi_0 \cdot \sin\epsilon\simeq 0.01\,\phi_0$ to the magnetic flux in the junction, leading to a nearly regular, but still inversely hysteretic Fraunhofer pattern with several side lobes in Fig.~\ref{InvHyst}(b).
We note an asymmetry in magnitude and width between the left and right secondary lobes with respect to the field sweep direction that is qualitatively comparable to the findings of Senapati~\emph{et~al.}~\cite{Senapati_2023}. Following the same protocol, sample~C is tested for hysteretic behavior under $B_z$, revealing similar inverted hysteresis, as shown in the Supplemental Material~\cite{SupplMat}.

In the following, we explain the observed inverted hysteresis based on the critical state model~\cite{Bean_1962}. This model considers strongly pinned vortices that gradually enter or exit a superconducting slab from the sides when the diamagnetic edge current exceeds the depinning current $j^\textrm{d}$ locally. Confined in vortices, the magnetic flux penetrates much deeper than the London penetration depth. According to Maxwell's equation $\nabla \times \,\vec{b} = \mu_0\,\vec{j}^\textrm{e}$, the gradual penetration of the local flux density $\vec{b}$ introduces an edge current density $\vec{j}^\textrm{e}$ in the outer regions of the superconductor until $|\vec{j}^\textrm{e}|$ equilibrates at $j^\textrm{d}$. These edge currents can largely exceed those present in the Meissner state.
According to Amp\'ere's law, the edge currents $\vec{j}^\textrm{e}$ generate a stray magnetic field $\vec{b}^\textrm{s}$. Figs.~\ref{InvHyst}(c-e) outline that the $z$-component $b^\textrm{s}_z$ can overcompensate the externally applied field $B_z$ in the junction area, causing an apparent inverted hysteresis in the flux. We depict a cross-section along the length of the junction in the $xz$-plane subject to an external field $B_z$. In this simplified geometry, Maxwell's equation reduces to
\begin{align} \label{eq:j^s}
    -\frac{\textrm{d}b_z(x)}{\textrm{d}x} = \mu_0 j_y^\textrm{e}(x)
\end{align}
such that the edge current $\vec{j}^\textrm{e}$ flows in $y$-direction.
The estimated local magnetic flux density $b_z(x) = B_z + b^\textrm{v}_z(x) + b^\textrm{s}_z(x)$, where $b^\textrm{v}_z(x)$ denotes the field contribution from the locally varying vortex density, is illustrated together with the edge current $j_y^\textrm{e}(x)$. Initially, a field $B_z^0 < B_\textrm{c2}$ is applied in positive $z$-direction, see Fig.~\ref{InvHyst}(c), where $B_\textrm{c2}$ is the upper critical field. By cooling below $T_\textrm{c}$ in fixed $B_z$, the Nb leads are brought into a fully penetrated Abrikosov state, such that $b_z(x) = B_z^0$ both inside the junction as well as inside the leads. No edge currents are induced.
Upon reducing the applied field below $B_z^0$, vortices start to be pushed out near the edges of the Nb strips such that a gradient of $b^\textrm{v}_z$ forms near the edges.
Thus we expect $j_y^\textrm{e}(x) \simeq \pm j^\textrm{d}$ in these regions as depicted in Fig.~\ref{InvHyst}(d). The resulting stray field $\vec{b}^\textrm{s}$ adds a negative $z$-component in the area of the junction. When $B_z=0$, the field in the junction area is $b_z = b^\textrm{s}_z < 0$ [see Fig.~\ref{InvHyst}(e)].
Consequently, the point of net zero flux enclosed in the junction must be reached while $B_z>0$ during a backward field sweep and $B_z<0$ during a forward sweep.
Taking $j^\textrm{d}\simeq 50\times 10^6\SI{}{A/cm^2}$ (see Refs.~\cite{Huebener_1975, Dedyu_1990, Park_1992, Pautrat_2004} and Supplemental Material~\cite{SupplMat}), the magnitude of the stray field at zero applied field can be estimated to be $\sim3$-$\SI{4}{mT}$ for the case of sample~B (see Supplemental Material~\cite{SupplMat}). Given the crudeness of the model, such values are consistent with the shifts of $\SI{\pm10}{mT}$ seen in Fig.~\ref{InvHyst}(a). Furthermore, the jumps in $I_\textrm{c}(B_z)$ visible in Figs.~\ref{oop} and~\ref{InvHyst}(a) as well as the minor loop behavior of sample~C can be explained with this model, as discussed in the Supplemental Material~\cite{SupplMat}. Also the inverted hysteresis observed under in-plane fields in Fig.~\ref{intro}(d) can be understood by the critical state model if we take into account a small residual out-of-plane component in the applied field. The contribution of each vortex to $b_z^\textrm{s}$ is strongly reduced due to the in-plane orientation. On the other hand, $B_\textrm{ip}$ in Figs.~\ref{intro}(d-f) is on the order of hundreds of mT, resulting in a much larger number of vortices compared to the case of Fig.~\ref{InvHyst}(a), thus producing a comparable hysteretic effect. Due to the small contribution of the pinned in-plane vortices to the stray field $b^\textrm{s}_z$, flux jumps cannot be resolved.

It should be emphasized that the mechanism outlined above is unrelated to the flux focusing effect that occurs in the Meissner state. The discussed edge currents are distinct from Meissner currents in that they arise from the inhomogeneous and hysteretic vortex distribution. Thus they can persist even when the external field is zero and give rise to an offset field in the junction which remains finite at zero applied field. Our observation is consistent with that of Ref.~\cite{Mandal_2022}.

\end{document}


\title{Supplemental Material for the article \\ Magneto-Chiral Anisotropy in Josephson Diode Effect of All-Metallic Lateral Junctions with Interfacial Rashba Spin-Orbit Coupling}

\author{Maximilian Mangold}
\affiliation{School of Natural Sciences, Technical University of Munich, 85748 Garching b. Munich, Germany}
\affiliation{Center for Quantum Engineering (ZQE), Technical University of Munich, 85748 Garching b. Munich, Germany}
\author{Lorenz Bauriedl}
\affiliation{Department of Physics, University of Regensburg, 93040 Regensburg, Germany}
\affiliation{Halle-Berlin-Regensburg Cluster of Excellence CCE, University of Regensburg, 93040 Regensburg, Germany}
\author{Johanna Berger}
\affiliation{Department of Physics, University of Regensburg, 93040 Regensburg, Germany}
\affiliation{Halle-Berlin-Regensburg Cluster of Excellence CCE, University of Regensburg, 93040 Regensburg, Germany}
\author{Chang Yu-Cheng}
\affiliation{Pico Group, QTF Centre of Excellence, Department of Applied Physics, Aalto University, P.O. Box 15100, FI-00076 Aalto, Finland}
\author{Thomas N.G. Meier}
\affiliation{School of Natural Sciences, Technical University of Munich, 85748 Garching b. Munich, Germany}
\affiliation{Center for Quantum Engineering (ZQE), Technical University of Munich, 85748 Garching b. Munich, Germany}
\author{Matthias Kronseder}
\affiliation{Department of Physics, University of Regensburg, 93040 Regensburg, Germany}
\affiliation{Halle-Berlin-Regensburg Cluster of Excellence CCE, University of Regensburg, 93040 Regensburg, Germany}
\author{Pertti Hakonen}
\affiliation{Low Temperature Laboratory, Department of Applied Physics, Aalto University, P.O. Box 15100, FI-00076 Espoo, Finland}
\author{Christian H. Back}
\affiliation{School of Natural Sciences, Technical University of Munich, 85748 Garching b. Munich, Germany}
\affiliation{Center for Quantum Engineering (ZQE), Technical University of Munich, 85748 Garching b. Munich, Germany}
\author{Christoph Strunk}
\affiliation{Department of Physics, University of Regensburg, 93040 Regensburg, Germany}
\affiliation{Halle-Berlin-Regensburg Cluster of Excellence CCE, University of Regensburg, 93040 Regensburg, Germany}
\author{Dhavala Suri}
\affiliation{School of Natural Sciences, Technical University of Munich, 85748 Garching b. Munich, Germany}
\affiliation{Centre for Nanoscience and Engineering, Indian Institute of Science, Bengaluru 560012, India}

\maketitle

\numsection{\\$I_\textrm{c}R_\textrm{n}$ product and Thouless energy of sample A}\label{SI:sec:C345_IcRn}

To estimate the $I_\textrm{c}R_\textrm{n}$ product of sample A, the maximum $I_\textrm{c}$ value of \SI{180}{\micro A} is taken from the main text Fig.~1(d), and $R_\textrm{n}\approx \SI{1}{\Omega}$ is inferred from the IV characteristics shown in Fig.~1(c). However, owing to the lead geometry of sample~A (see SEM image in the main text Fig.~1(b)), $I_\textrm{c}$ is governed by the narrowest section of the junction (encircled in the SEM image) whereas the measured value of $R_\textrm{n}$ is given also by the remaining metallic film between the Nb leads which contributes to the total conductance in the normal state of the junction. This additional area is \SI{4}{\micro m} wide, compared to a width of \SI{180}{nm} of the narrow part. Assuming a constant conductivity in the Fe/Al/Pt film, the narrow region can be estimated to contribute roughly $\SI{180}{nm}/\SI{4}{\micro m} \approx 5\%$ of the total normal state-conductance. Hence, to compare to the $I_\textrm{c}$ value, the measured value of $R_\textrm{n}$ has to be increased by a factor of $\approx 1/0.05$, thus $I_\textrm{c}R_\textrm{n} = \SI{3.0}{mV}$ for sample A. Using a superconducting gap for the Nb leads of $\Delta_\textrm{Nb} = 1.764 k_\textrm{B}T_\textrm{c} = \SI{1.3}{meV}$ with $T_\textrm{c} = \SI{8.6}{K}$, we find $eI_\textrm{c}R_\textrm{n} / \Delta_\textrm{Nb} \approx 2.3$.

The Thouless energy can be estimated to $E_\textrm{Th} = \hbar D_\textrm{Pt} / l^2 \approx \SI{1.6}{meV} = 1.2 \Delta_\textrm{Nb}$ using $D_\textrm{Pt} = \SI{0.001}{m^2/s}$ and $l = 20 \pm \SI{5}{nm}$. Thus, $eI_\textrm{c}R_\textrm{n} / E_\textrm{Th} \approx 1.9$. Given the uncertainties, we conclude that the junction is in the intermediate regime between the short and the long junction limits~\cite{Dubos_2001}.

\newpage

\numsection{\\Critical current corresponding to the $\eta$ vs $B_\textrm{ip}$}\label{SI:sec:C345_rawIcs}

\begin{figure}[tbh]
\centering
\includegraphics[width=5in]{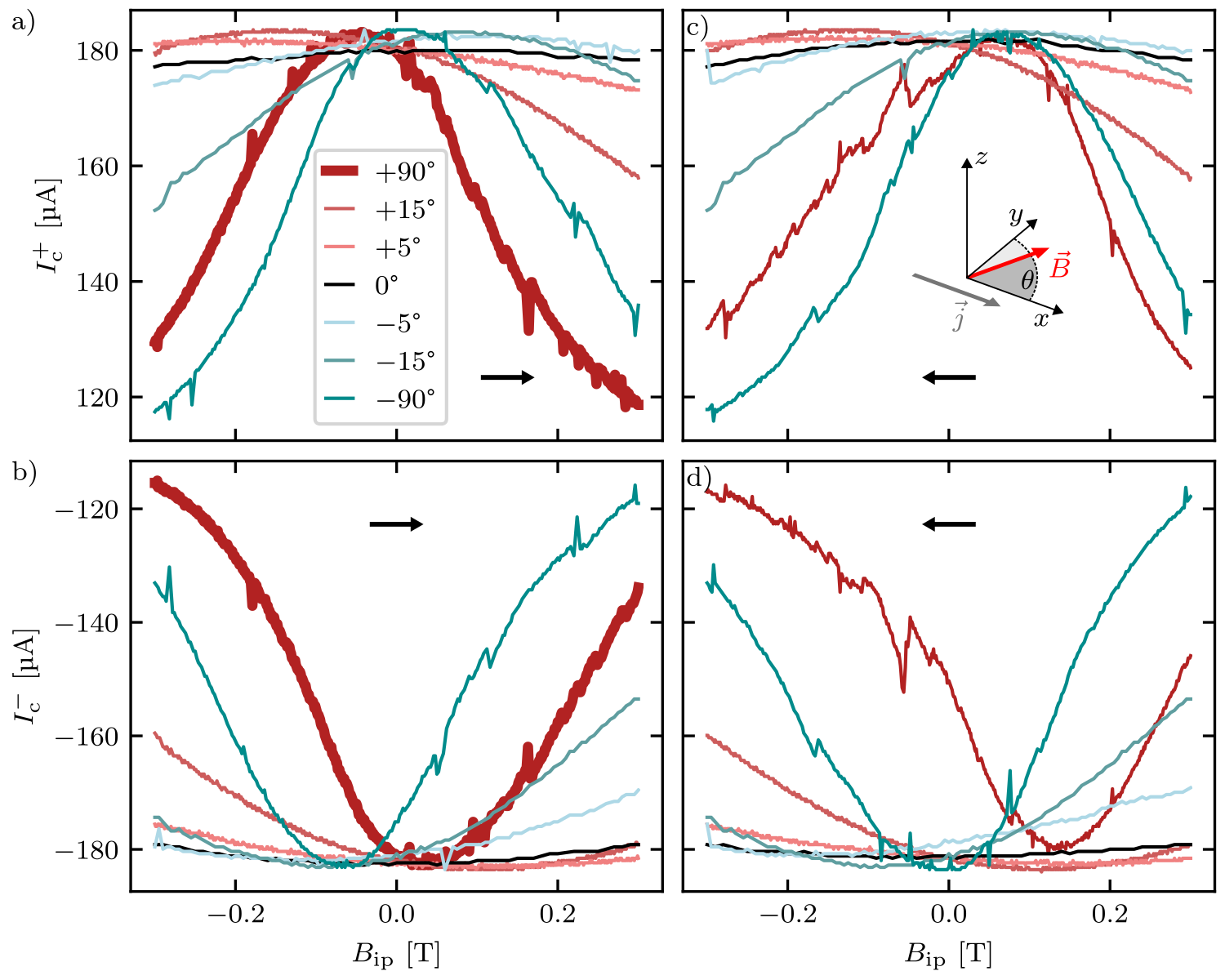}
\caption{Critical current of sample A under in-plane magnetic field for various in-plane field angles $\theta$ under (a) positive bias and forward field sweep, (b) negative bias and forward field sweep, (c) positive bias and backward field sweep, and (d) negative bias and backward field sweep. Inset in panel (c) indicates field direction with respect to applied current and sample $xy$-plane.}
\label{SI:fig:C345_rawIcs}
\end{figure}

Fig.~\ref{SI:fig:C345_rawIcs} corresponds to the critical current curves which generate $\eta$ in Fig.~1(e) in the main text.

The diode factor $\eta$ discussed in the main text is calculated from measurements of $I_\textrm{c}^+$ and $I_\textrm{c}^-$ in forward field sweep direction, plotted in Figs.~\ref{SI:fig:C345_rawIcs}(a) and (b), respectively. The forward sweep at $\theta=\SI{+90}{^\circ}$ is highlighted to visualize the diode effect, manifested in a horizontal shift between $I_\textrm{c}^+$ and $I_\textrm{c}^-$. Figs.~\ref{SI:fig:C345_rawIcs}(c) and (d) show $I_\textrm{c}^+$ and $I_\textrm{c}^-$ in backward sweep direction, also revealing a diode effect. Inverted hysteresis is observed at all angles $\theta$.

\newpage

\numsection{\\Details on sample B}\label{SI:sec:CuPt}

\begin{figure*}[tbh]
\centering
\includegraphics[width=\linewidth]{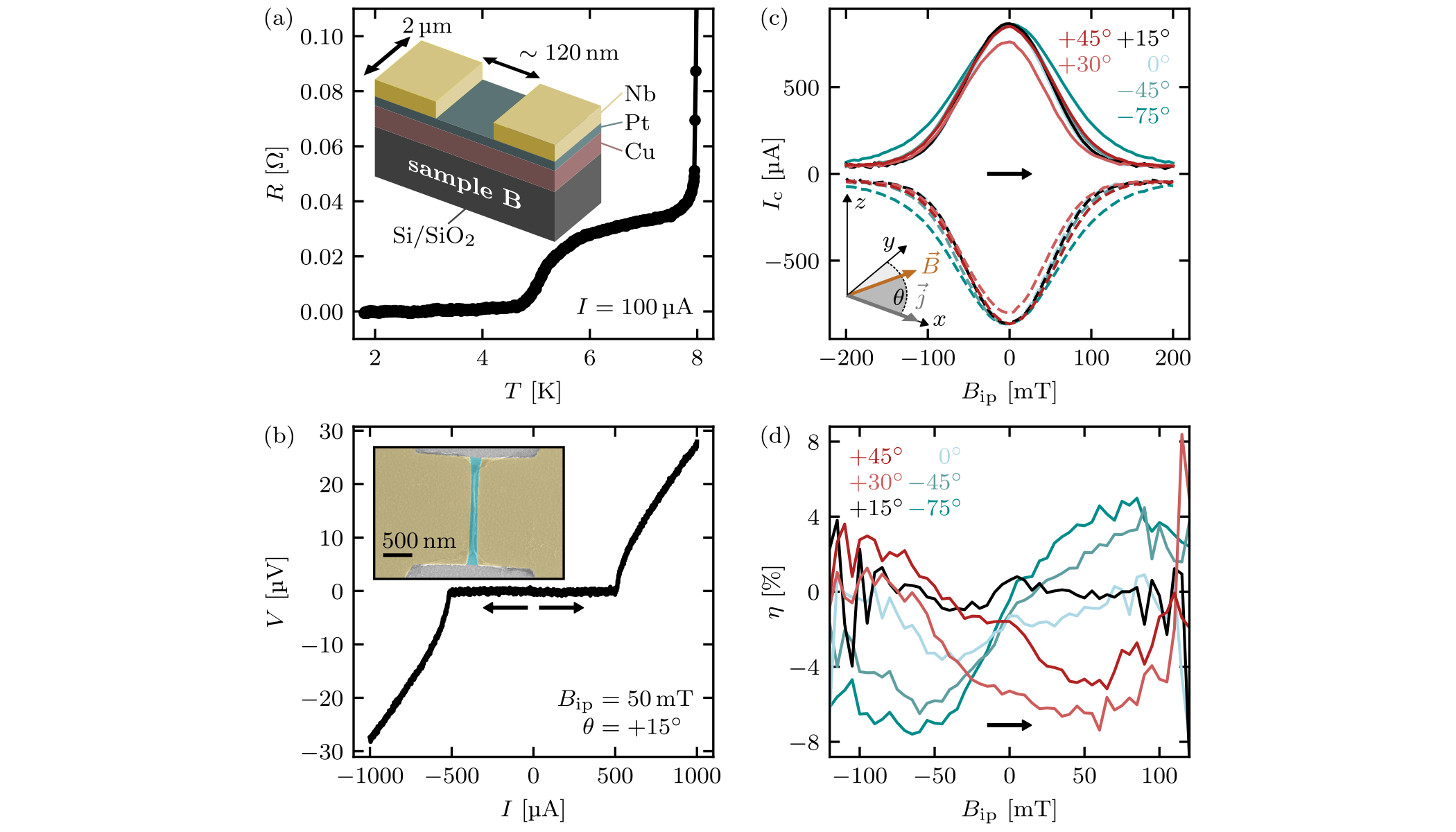}
\caption{Details and raw data on sample B. (a) Temperature-dependent resistance measured at \SI{100}{\micro A} in zero magnetic field. A constant zero-bias voltage is subtracted. Inset: Schematic diagram of the Cu/Pt/Nb stack and the JJ on top of a Si/SiO$_2$ substrate. (b) Typical IV curve, acquired by sweeping the applied current from 0 to $\pm\SI{1000}{\micro A}$. The inset shows a false-color SEM image of the device where the Nb leads are highlighted in yellow and the junction area is colored in blue. (c) $I_\textrm{c}(B_\textrm{ip})$ at various in-plane angles $\theta$. The field orientation is indicated by the inset. (d) Diode factor extracted from the data shown in panel (c) before shifting the data at $\theta=+30^\circ$ and $+15^\circ$ to generate the main text Fig.~1(f).}
\label{SI:fig:CuPt}
\end{figure*}

The resistance of sample~B as a function of temperature is shown in Fig.~\ref{SI:fig:CuPt}(a). It exhibits a sharp edge around $T=\SI{8.0}{K}$, indicating the superconducting transition of the Nb leads, and a second, broader transition near \SI{5.0}{K} that is attributed to the proximitized junction area. Above \SI{8.0}{K}, $R = \SI{1.3}{\Omega}$. Below \SI{5.0}{K}, IV characteristics are measured by sweeping the applied current from zero to \SI{+1000}{\micro A} and from zero to \SI{-1000}{\micro A}. Fig.~\ref{SI:fig:CuPt}(b) shows a typical IV curve measured at $B_\textrm{ip} = \SI{0.05}{T}$ and $\theta = \SI{+15}{^\circ}$, revealing $I_\textrm{c}R_\textrm{n} \approx \SI{26}{\micro V}$, which is comparable to literature values for similar Cu weak links~\cite{Senapati_2023, Garcia_2009}. 

Fig.~\ref{SI:fig:CuPt}(c) depicts the field-dependence of $I_\textrm{c}(B_\textrm{ip})$ of sample~B for several in-plane field angles in forward field sweep direction, showing regular, $\sim \SI{200}{mT}$ wide single lobes. The diode factor $\eta$ extracted from this data is shown in Fig.~\ref{SI:fig:CuPt}(d). While the $\theta$-dependence is less clear than in the case of sample A, a polarity change between $\theta=\SI{+90}{^\circ}$ and \SI{-90}{^\circ} is observed. To generate the main text Fig.~1(f), a zero-field offset of \SI{-3}{\%} and $+\SI{1.5}{\%}$ is subtracted from the curves measured at $\theta = +30^\circ$ and $+15^\circ$, respectively, to align the curves at $B_\textrm{ip}=0$. The offset is considered a random fluctuation, and subtracting it does not conflict with the interpretation of the diode factor as being related to Rashba SOC.

\newpage

\numsection{\\Fabrication and characterization of sample C}\label{SI:sec:Cu-only}

\begin{figure*}[tbh]
\centering
\includegraphics[width=\linewidth]{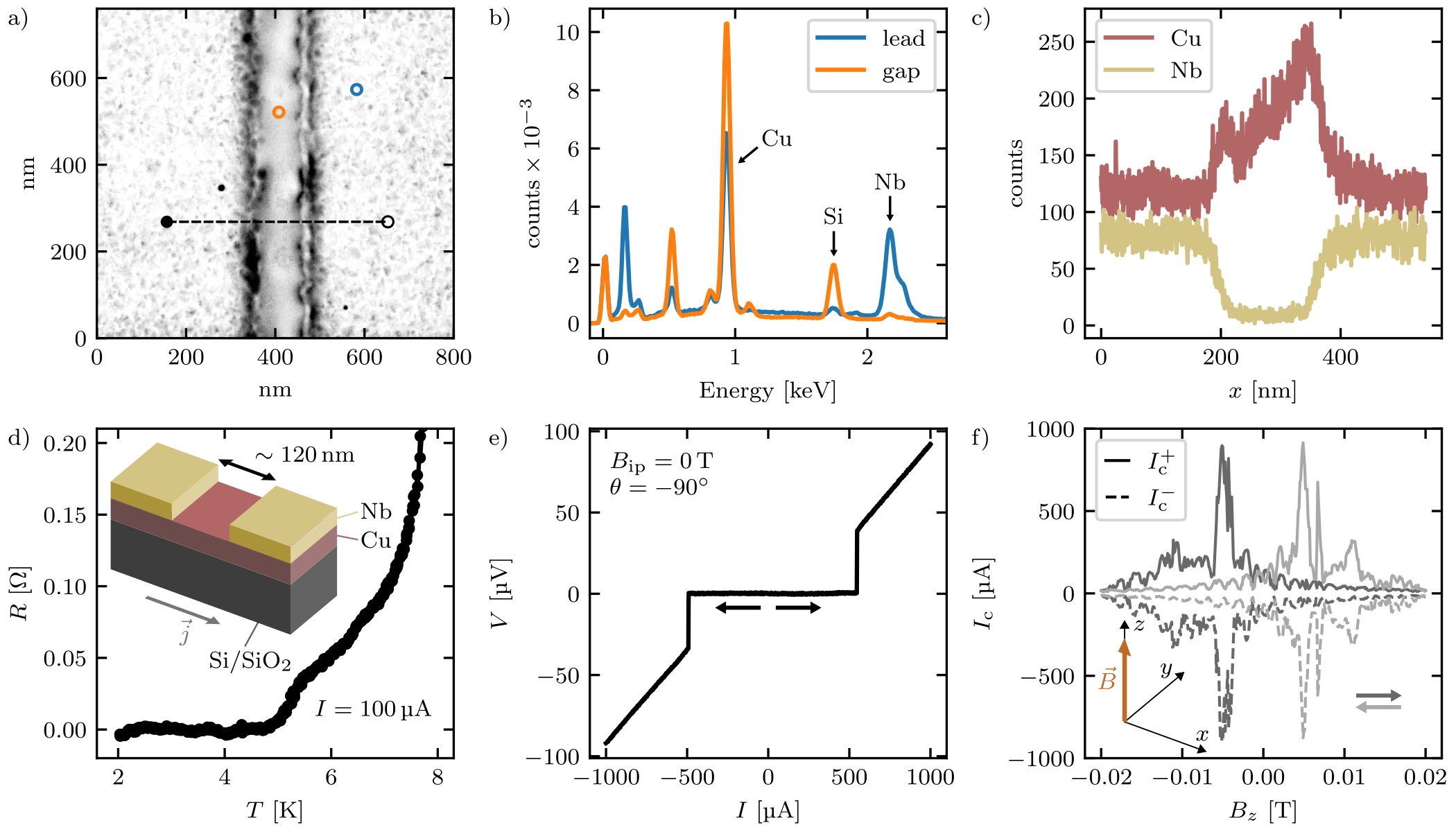}
\caption{Fabrication and characterization of sample C. (a) SEM image of central section of the junction. The colored circles mark the positions of the EDX spectra shown in panel (b). The dashed line indicates the path of the EDX line scan depicted in panel (c). (b) EDX spectra taken on the Nb leads and inside the Cu junction at an acceleration voltage of \SI{5}{kV}. The spectral peaks of Cu, Si and Nb are indicated. (c) Line scan of element-specific counts across the length of the junction. (d) $R(T)$ curve, measured at $I = \SI{100}{\micro A}$ and $B=\SI{0}{T}$. The inset shows the schematic device structure and the current direction. (e) IV characteristics measured at $B_\textrm{ip}=\SI{0}{T}$ and $\theta=-90^\circ$. The sweeping direction of the applied current is indicated by the arrows. (f) $I_\textrm{c}(B_z)$ in forward and backward field sweep direction, as indicated by the arrows.}
\label{SI:fig:Cu-only}
\end{figure*}

During the fabrication on the GaAs/Fe/Al/Pt/Nb and Cu/Pt/Nb stacks, electron beam lithography-defined junctions were etched into the Nb by SF$_6$ reactive ion etching, using the Pt layer as an etch stop. For the Cu/Nb sample, selective etching becomes more difficult due to the high etch rate of Cu in an SF$_6$ environment. To precisely control the etch depth, a Ga-based Focussed Ion Beam microscope (Zeiss Crossbeam 550) is employed at an acceleration voltage of \SI{30}{kV} and a beam current of \SI{1}{pA}. The etch rate is controlled to stop inside the Cu layer. SEM and Electron Dispersive X-ray spectroscopy (EDX) are applied in-situ immediately after the etch step. An SEM image of the central region of the Cu junction is depicted in Fig.~\ref{SI:fig:Cu-only}(a). EDX point spectra are taken on the Nb leads and inside the junction and shown in Fig.~\ref{SI:fig:Cu-only}(b) in blue and orange, respectively, as indicated by the colored circles in Fig.~\ref{SI:fig:Cu-only}(a). Elementary spectra are fitted to the curves using the Oxford Instruments AZtec software tool. The characteristic peaks of relevant elements are highlighted in Fig.~\ref{SI:fig:Cu-only} (b). Compared to the full stack on the leads, the Cu and Si peaks are enhanced inside the gap, while the Nb peak is largely suppressed.
The small remaining Nb component inside the gap is attributed to unavoidable re-deposition during the etch process; even after etching deep into the Cu layer, the Nb peak does not fully disappear (not shown). In Fig.~\ref{SI:fig:Cu-only}(c), the Cu and Nb parts of a line scan spectrum across the junction are compared. Inside the junction, the Nb part reduces to almost zero while the Cu counts increase. In Fig.~\ref{SI:fig:Cu-only}(d), the temperature-dependent resistance of sample C shows a steep decline when cooling below \SI{8}{K} as the Nb leads turn superconducting. Around \SI{5}{K}, a second step is observed, attributed to the junction entering the superconducting phase. A typical IV curve measured at $\theta=-90^\circ$ and zero field is shown in Fig.~\ref{SI:fig:Cu-only}(e). The magneto-interference pattern under $B_z$ in forward and backward sweep direction is depicted in Fig.~\ref{SI:fig:Cu-only}(f), demonstrating inverted hysteresis of $\sim \SI{10}{mT}$, comparable to sample B discussed in the main text. Spontaneous jumps in $I_\textrm{c}$ are observed, as discussed further in Supplemental Note~8. For both sweep directions, the trailing side of the spectrum after reaching maximum $I_\textrm{c}$ exhibits a periodic Fraunhofer pattern which shows the Josephson junction character of the device.

\newpage

\numsection{\\Comparison of in-plane diode effect between GaAs/Fe/Al/Pt, Cu/Pt and Cu weak links}\label{SI:sec:fishes}

\begin{figure}[tbh]
\centering
\includegraphics[width=\linewidth]{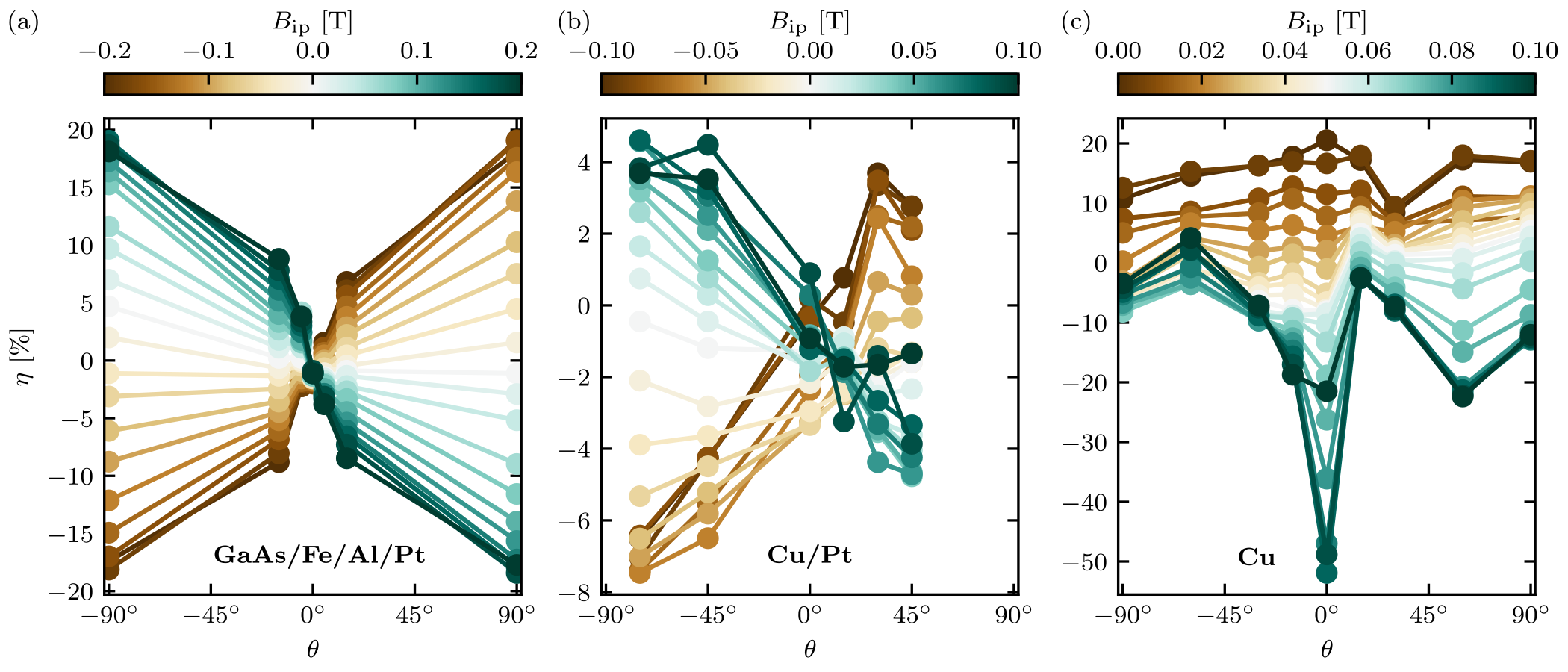}
\caption{Comparison of in-plane field-dependent diode effect of all samples presented in the main text. $\eta(\theta)$ is shown for (a) sample A, (b) sample B and (c) sample C at several magnetic field values. The data for devices A and B corresponds directly to the curves shown in main text Fig.~1(e) and (f), respectively.}
\label{SI:fig:fishes}
\end{figure}

The in-plane diode effect of the three measured samples with weak links consisting of GaAs/Fe/Al/Pt (sample A), Cu/Pt (sample B) and Cu (sample C), is summarized in Fig.~\ref{SI:fig:fishes}. To illustrate the fundamental difference between the devices A and B on one side and sample C on the other, $\eta$ is plotted for various in-plane fields as a function of the in-plane angle $\theta$. This corresponds to taking vertical line cuts in the $\eta(B_\textrm{ip})$ curves shown in the main text Fig.~1(e) and (f) for devices A and B, respectively. For sample A (Fig.~\ref{SI:fig:fishes} (a)), a sign change in $\eta$ occurs at $\theta=0^\circ$. Similarly, in sample~B (Fig.~\ref{SI:fig:fishes}(b)), a sign change is observed between zero and $20^\circ$. This offset angle is likely related to a misalignment between the nominal field direction and the current direction on the chip. For sample C, however, no change of polarity can be identified in Fig.~\ref{SI:fig:fishes}(c). For an individual field, the diode factor is finite, but displays an even symmetry in $\theta$. Thus, the diode effect in sample~C does not comply with the symmetry of Rashba SOC.

\newpage

\numsection{\\Time-reversal symmetry including magnetic field sweep direction}\label{SI:sec:TRS}

\begin{figure}[tbh]
\centering
\includegraphics[width=\linewidth]{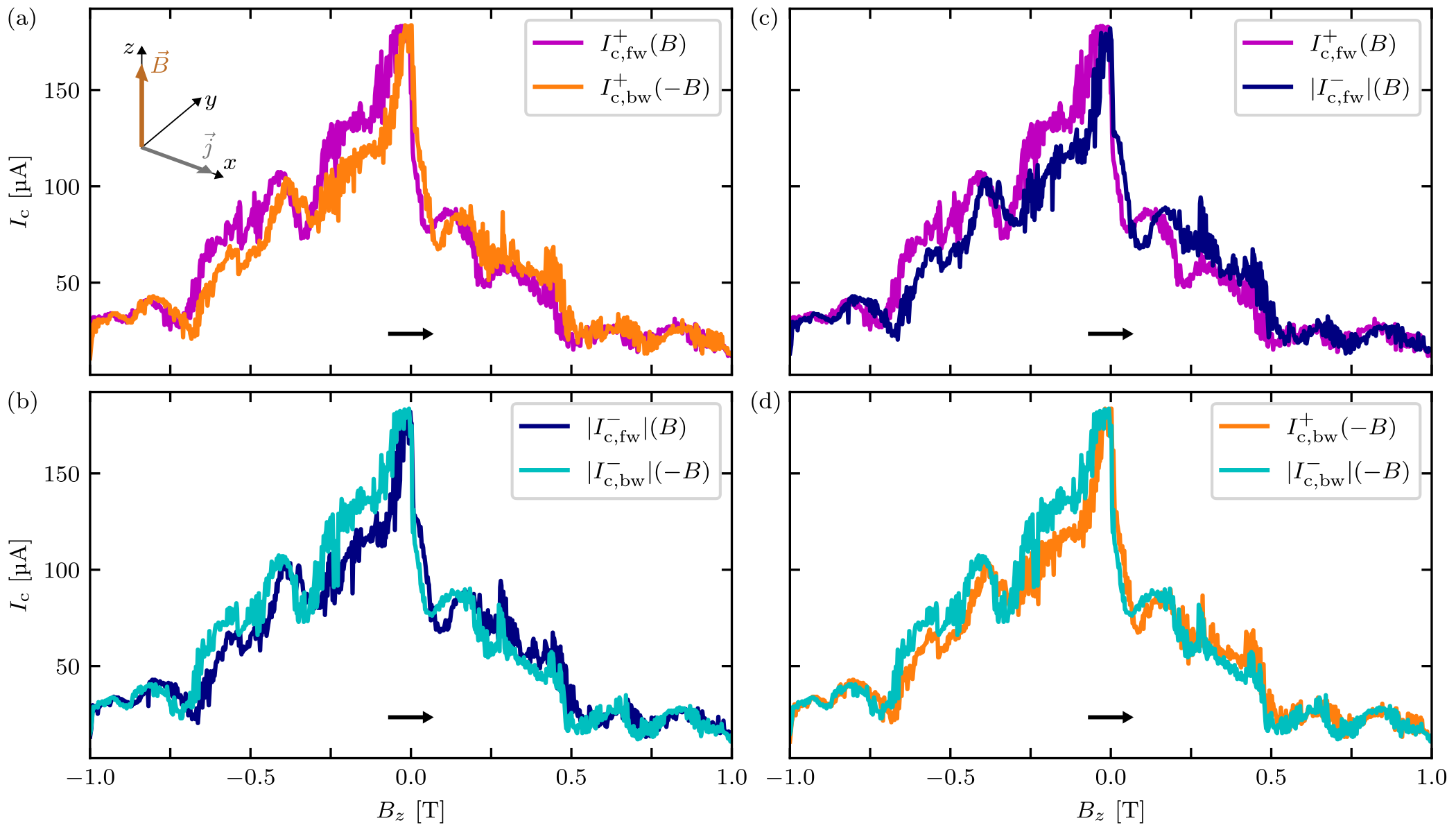}
\caption{Comparison of $I_\textrm{c}$ curves for sample A under $B_z$ in both field sweep directions without consideration of the combined reversal of current, field and field sweep direction. The data is reproduced from the plots shown in Fig.~2 in the main text. The field axis of the backward sweep data is flipped to match the forward sweep direction. In (a) and (b), curves of opposite field and field sweep direction are compared. In (c) and (d), curves at opposite current and field are compared.}
\label{SI:fig:TRS}
\end{figure}

It is shown in the main text Fig.~2 that the combined reversal of current, field and field sweep direction matches different parts of the $I_\textrm{c}(B_z)$ curve of sample~A. Here, we demonstrate that neglecting one of those operations does not lead to an adequate overlap. Fig.~\ref{SI:fig:TRS}(a) shows $I_\textrm{c,fw}^+(B_z)$ and $I_\textrm{c,bw}^+(-B_z)$ of sample A corresponding to Fig.~2. For the latter, the field axis is reversed to account for the backward sweep direction. In this combination, field and field sweep direction are reversed while the current direction is maintained. A significant mismatch in the prominent features is found between the two curves. An equivalent comparison is made in Fig.~\ref{SI:fig:TRS}(b) between $|I_\textrm{c,fw}^-|(B_z)$ and $|I_\textrm{c,bw}^-|(-B_z)$, leading to a similar mismatch. Alternatively, combining $I_\textrm{c,fw}^+(B_z)$ and $|I_\textrm{c,fw}^-|(B_z)$, as shown in Fig.~\ref{SI:fig:TRS}(c), corresponds to the reversal of current and field direction without inverting the field sweep direction. Again, no adequate overlap is achieved. Similarly, Fig.~\ref{SI:fig:TRS}(d) demonstrates the mismatch between $I_\textrm{c,bw}^+(-B_z)$ and $|I_\textrm{c,bw}^-|(-B_z)$. Therefore, the reversal of the field sweep direction must be included in the time reversal operations $I_\textrm{c}^+ \rightarrow I_\textrm{c}^-$ and $+B_z \rightarrow -B_z$.

\begin{figure*}[tbh]
\centering
\includegraphics[width=\linewidth]{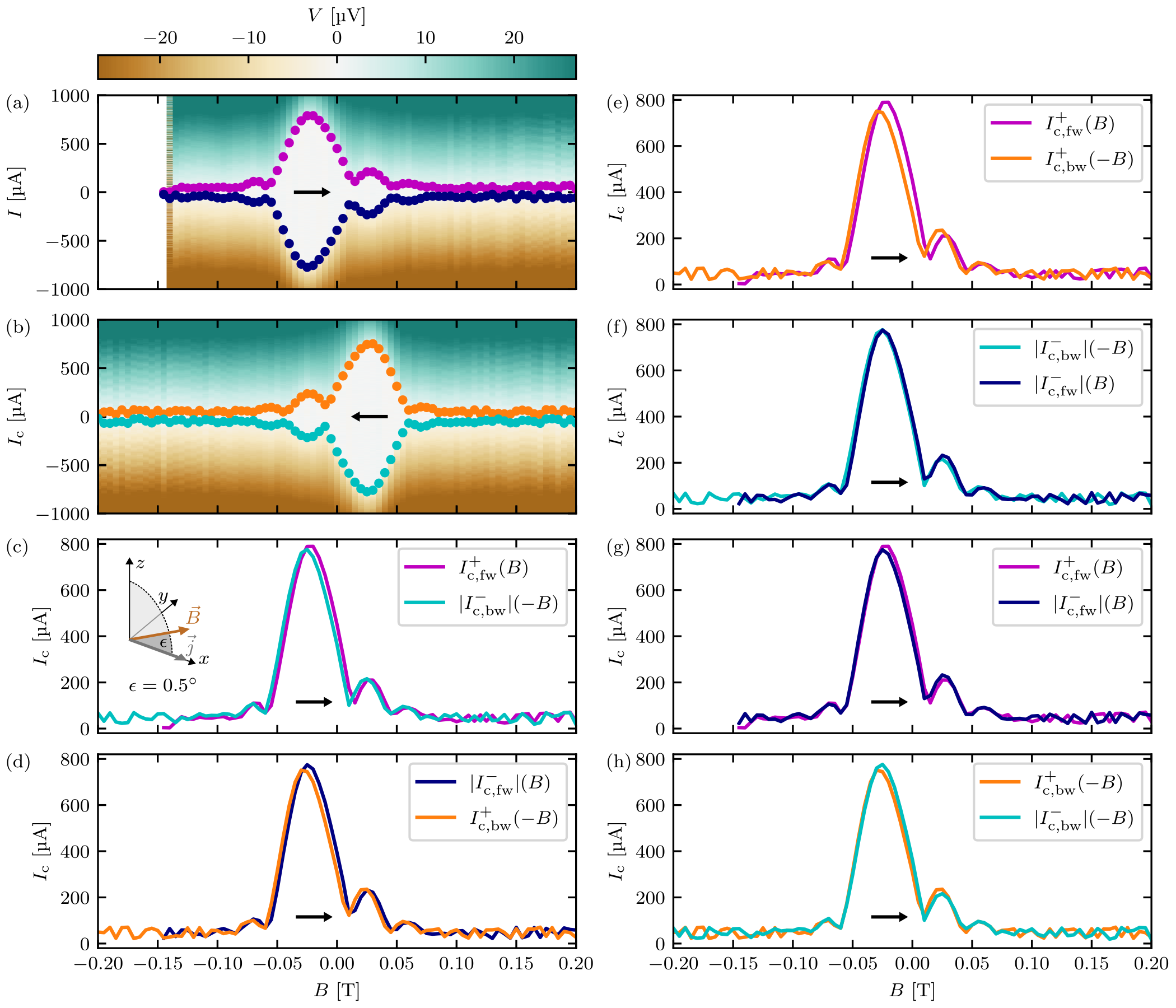}
\caption{TRS partners for small-tilt out-of-plane fields in sample B. (a) and (b) show the full field-dependent IV color map for $\theta=0^\circ$ with a small misalignment angle $\epsilon=0.5^\circ$ from the sample plane. The extracted critical current corresponds to the data shown in the main text Fig.~3(b). Panels (c) and (d) match the $I_\textrm{c}$ curves with the respective TRS partner under consideration of the field sweep direction (analogue to the main text Fig.~2(c) and (d) in sample A). Panels (e-h) overlay the same data in other combinations.}
\label{SI:fig:TRSmisaligned}
\end{figure*}

The overlap of $I_\textrm{c}$ curves between forward and backward field sweep direction is also investigated for the sample B. Due to the jumps in $I_\textrm{c}$ discussed in Supplemental Note~8, the matching of $I_\textrm{c}$ curves under time reversal operations is not possible in the case of a perpendicular field. When applying the field close to the sample plane with only a small out-of-plane misalignment angle $\epsilon$, vortices are pinned across the full width of the Nb leads, leading to a stable Fraunhofer pattern, as shown in the main text Fig.~3(b). Figs.~\ref{SI:fig:TRSmisaligned}(a) and (b) depict the corresponding IV color maps in both field sweep directions. The comparison of $I_\textrm{c}^+$ and $|I_\textrm{c}^-|$ under the application of time reversal operations including the sweep directions is given in Figs.~\ref{SI:fig:TRSmisaligned}(c) and (d) while the other combinations that neglect one of the symmetry operations are shown in Figs.~\ref{SI:fig:TRSmisaligned}(e-h). In this case, no significant enhancement of overlap can be identified when including the sweep direction reversal.

\newpage

\numsection{\\Details on the inverted hysteresis of sample B}\label{SI:sec:CuPt_invhyst-demonstrations}

\begin{figure*}[tbh]
\centering
\includegraphics[width=\linewidth]{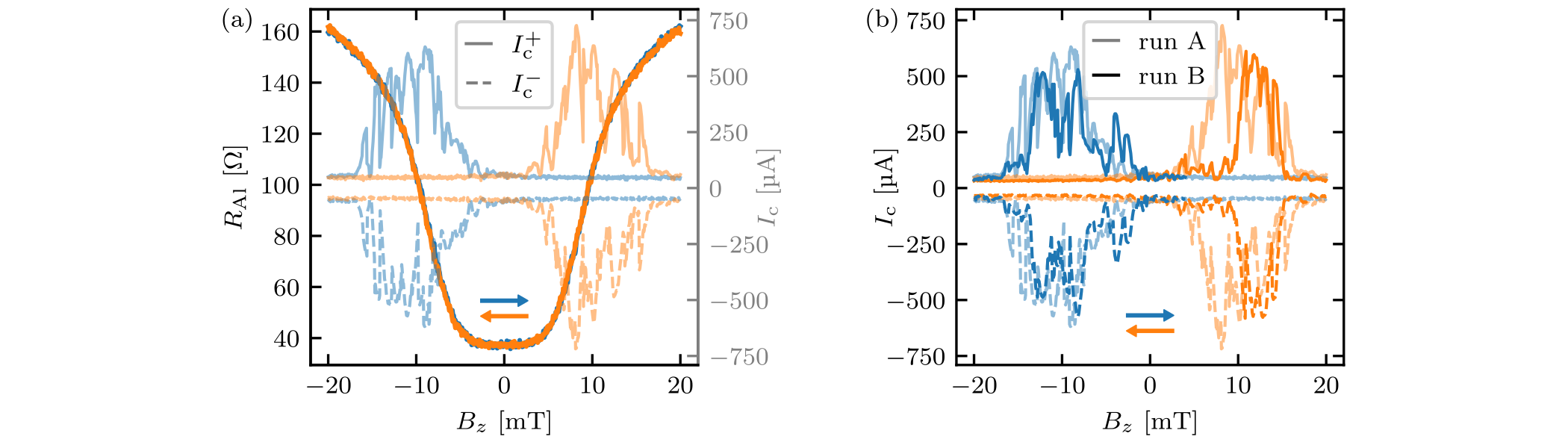}
\caption{Details on the inverted hysteresis of sample~B. (a) Right axis: $I_\textrm{c}(B_z)$, re-plotted from the main text Fig.~3(a). Left axis: Al stripe resistance, measured simultaneously with $I_\textrm{c}(B_z)$. The arrows indicate the field sweep direction. (b) $I_\textrm{c}(B_z)$ measured on the same device in two different runs at \SI{1.8}{K}.}
\label{SI:fig:CuPt_invhyst-demonstrations}
\end{figure*}

Fig.~\ref{SI:fig:CuPt_invhyst-demonstrations}(a) depicts $I_\textrm{c}(B_z)$ of sample~B, also shown in the main text Fig.~3(a), in forward and backward magnetic field sweep direction, showing inverted hysteresis. In addition, the Al stripe resistance measured simultaneously (see Supplemental Note~11) is shown for both sweep directions in the corresponding colors. No hysteretic shift is observed in $R_\textrm{Al}$, demonstrating that the inverted hysteresis in $I_\textrm{c}(B_z)$ is not an experimental artifact.

The reproducibility of the $I_\textrm{c}(B_z)$ curves is investigated by comparing measurements of $I_\textrm{c}(B_z)$ from different runs, as shown in Fig.~\ref{SI:fig:CuPt_invhyst-demonstrations}(b). Run A corresponds to the data plotted in the main text Fig.~3(a), run B is recorded separately on the same device. While inverted hysteresis of similar width is present in both experiments, the jumps in $I_\textrm{c}$ occur at random fields. This indicates spontaneous re-ordering of the vortex distribution near the lead edges, leading to jumps in the flux through the junction, see also Supplemental Note~8.

\newpage

\numsection{\\Explanation of jumps in $I_\textrm{c}(B_z)$ and minor loops in $B_z$ through critical state model}\label{SI:sec:BeanConsequences}

\begin{figure*}[tbh]
\centering
\includegraphics[width=\linewidth]{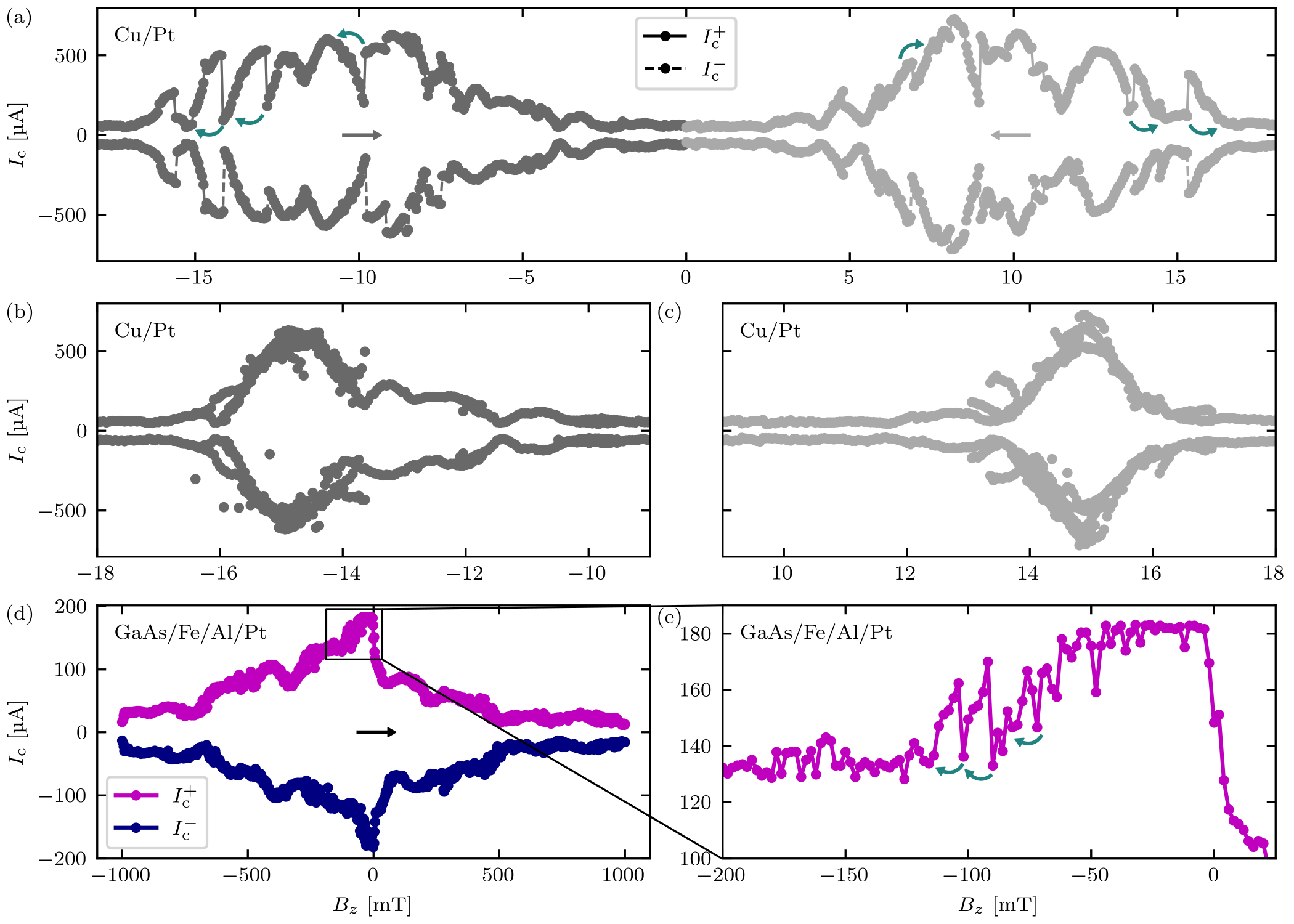}
\caption{Random jumps in $I_\textrm{c}(B_z)$ under consideration of critical state model. (a) $I_\textrm{c}(B_z)$ of sample B in both sweep directions. The data is the same as shown in the main text Fig.~3(a). The blue arrows indicate examples of backward shifts with respect to the field sweep direction. (b) and (c) show $I_\textrm{c}(B_z)$ of sample B in forward and backward field sweep direction, respectively, after performing the backward shifts. (d) $I_\textrm{c}(B_z)$ of sample A in forward field sweep direction, as shown in the main text Fig.~2(a). (e) Section of $I_\textrm{c}^+$ near zero field, as indicated in panel (d). The blue arrows indicate examples of backward shifts.}
\label{SI:fig:RandomJumps}
\end{figure*}

The jumps in $I_\textrm{c}$ observed for devices B and C under a perpendicular field $B_z$ can be understood when considering the critical state model (see main text) in combination with spontaneous reordering of the vortex distribution. Fig.~\ref{SI:fig:RandomJumps}(a) depicts $I_\textrm{c}(B_z)$ for sample B. The jumps in $I_\textrm{c}$ occur at random fields, but simultaneously in $I_\textrm{c}^+$ and $I_\textrm{c}^-$. Furthermore, the data following a jump can be shifted backwards with respect to the field sweep direction to overlap with the data before the jump. The blue arrows in Fig.~\ref{SI:fig:RandomJumps}(a) indicate examples of such backward shifts. By applying this method to both forward and backward field sweeps, irregular, but mostly consistent Fraunhofer patterns can be restored, as shown in Fig.~\ref{SI:fig:RandomJumps}(b) and (c), respectively. It is important to highlight that the data is only shifted opposed to the sweep direction and never along it. This behavior can be explained based on vortices inside the Nb leads. Following the critical state model as outlined in the main text, an ordered vortex distribution builds up during the field sweep which creates an additional stray field component on the junction area. If changes in the applied field spontaneously bring this distribution to a less ordered state through the random depinning of vortices, the stray field is reduced. Consequently, the total field acting on the junction is increased again (\emph{i.e.} more negative (positive) for the forward (backward) sweep before reaching zero) and a part of the junction's interference pattern is repeated at a lower applied field. In other words, the sections of the raw data in Fig.~\ref{SI:fig:RandomJumps}(a) constitute replica of a consistent Fraunhofer patterns. Therefore, shifting the curve backwards at each jump in $I_\textrm{c}$ reconstructs the intrinsic diffraction pattern of the junction. Quantitative deviations between the individual sections may be caused by the local inhomogeneity of the vortex distribution.

Since the Nb leads of devices A and B are of comparable thickness and width, a similar effect is expected in $I_\textrm{c}(B_z)$ of sample A. Fig.~\ref{SI:fig:RandomJumps}(d) shows the same $I_\textrm{c}$ curves as presented in the main text Fig.~2(a). A section of $I_\textrm{c}^+$ near $B_z=\SI{0}{mT}$ is shown in Fig.~\ref{SI:fig:RandomJumps}(e). At the small field scale, several jumps similar to those observed in Fig.~\ref{SI:fig:RandomJumps}(a) can be identified. We conclude that, due to the smaller junction size in sample A, the effect of a random reordering of the vortex distribution inside the leads is concealed by the $\sim \SI{1}{T}$ broad intrinsic interference pattern. Therefore, $I_\textrm{c}(B_z)$ is stable and reproducible, as opposed to the devices B and C where the flux change originating from vortex reordering is comparable to the width of the Fraunhofer pattern.

\newpage

\numsection{\\Estimation of the magnitude of the stray field inside the junction}\label{SI:sec:BscreenEstimation}

\begin{figure*}[tbh]
\centering
\includegraphics[width=\linewidth]{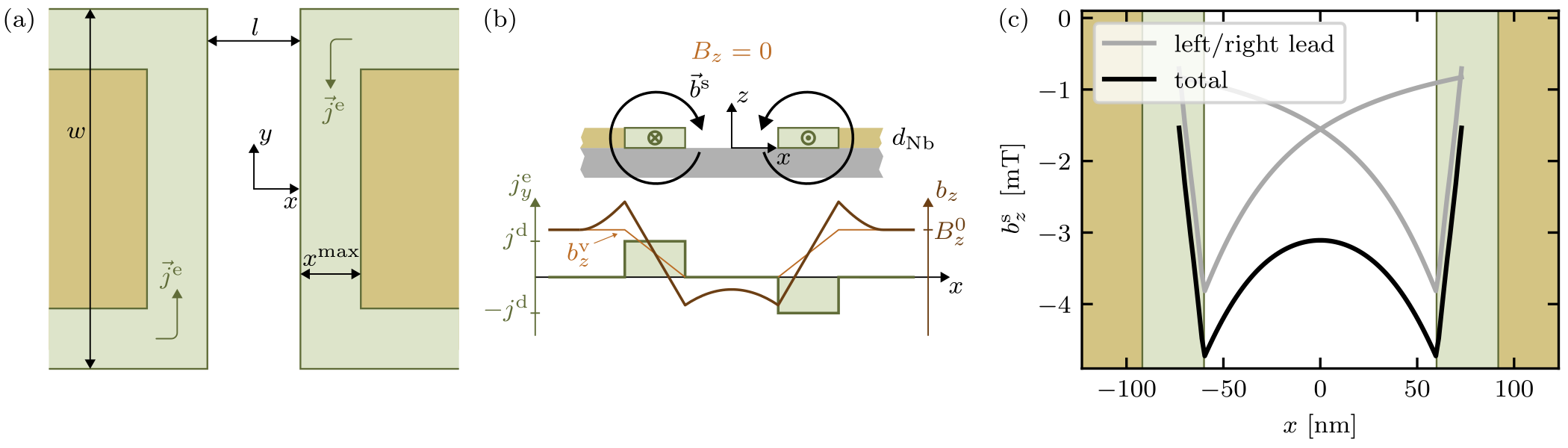}
\caption{Quantitative estimation of the inverted hysteresis from the critical state model. (a)~Top-view sketch of a junction of length $l$ with Nb leads of width $w$ carrying the edge currents~$\vec{j}^\textrm{e}$ down to a depth $x^\textrm{max}$, following from the critical state model. Origin and orientation are indicated by the coordinate system. (b) Side-view sketch of local field density $b_z$, vortex field $\vec{b}^\textrm{v}$, stray field $\vec{b}^\textrm{s}$ and edge current density $j_y^\textrm{e}$ after decreasing the applied magnetic field $B_z$ from $B_z^0$ to zero, analogue to the main text Fig.~3(e). (c) Result of the numerical integration following Eq.~(\ref{SI:eq:NumInt}) for the $z$-component of the stray field $b_z^\textrm{s}$ for the situation depicted in panel (b).}
\label{SI:fig:BscreenEstimation}
\end{figure*}

The explanation of inverted hysteresis of the magnetic field interference patterns on all presented junctions based on the critical state model is discussed qualitatively in the main text. Using a set of crude assumptions, the magnitude of the magnetic field $b^\textrm{s}$ produced by edge currents inside the lead edges acting on the junction area can be estimated. In the following, numerical values of sample~B are used for reference.

First, we address the vortex depinning current density $j^\textrm{d}$ which ultimately defines the penetration depth of the edge currents into the leads. The leads have a width of $w = \SI{2}{\micro m}$, as depicted in Fig.~\ref{SI:fig:BscreenEstimation}(a). Combined with the Nb thickness $d_\textrm{Nb} = \SI{40}{nm}$ and a critical current of $I_\textrm{c}^\textrm{leads}=\SI{40}{mA}$ determined on a different device, the critical depinning current density is $j^\textrm{d} = \frac{I_\textrm{c}^\textrm{leads}}{w ~ d_\textrm{Nb}}$. Further assuming $j^\textrm{d}$ to be constant everywhere inside the leads, the edge current depth $x^\textrm{max}$ of the vortices after ramping the applied field from $B_z^0$ to zero can be calculated, following Eq.~(1) of the main text:

\begin{align}\label{SI:eq:xmax}
-\frac{\textrm{d}b_z(x)}{\textrm{d}x} = -\frac{B_z^0}{x^\textrm{max}} = \mu_0 j_y^\textrm{e}(x) = \mu_0 j^\textrm{d}.
\end{align}

Using $B_z^0=\SI{20}{mT}$, the applied out-of-plane field at which the sample is field-cooled for the measurements shown in main text Fig.~3(a), one obtains $x^\textrm{max} = \SI{32}{nm}$. This defines a volume $x^\textrm{max} \times w \times d_\textrm{Nb}$ in which a constant current density $j^\textrm{d}$ flows in along the lead edges as depicted in Figs.~\ref{SI:fig:BscreenEstimation}(a) and (b). The $z$-component of the magnetic field generated by $j_y^\textrm{e}$ of a single lead can be calculated (along the $x$-axis) based on Biot-Savart's law by

\begin{align}\label{SI:eq:NumInt}
b_{z}^{\textrm{s,}i}(x) = \frac{\mu_0 j^\textrm{d}}{4\pi} \int_{-l/2-x^\textrm{max}}^{-l/2} \int_{-w/2}^{w/2} \int_{0}^{d_\textrm{Nb}} \frac{x-x'}{\left[ (x-x')^2 + y'^2 + z'^2 \right]^{3/2}} ~ \textrm{d}z' ~ \textrm{d}y' ~ \textrm{d}x' 
\end{align}

where the index $i$ denotes the individual leads. Fig.~\ref{SI:fig:BscreenEstimation}(c) shows the result of a numeric integration applied for each lead, and the total stray field $b_z^\textrm{s}(x)$. The result implies that at zero applied field, the junction is subject to a field density of $3\mbox{-}\SI{4}{mT}$ which is comparable to the inverted hysteresis width of $\pm \SI{10}{mT}$ observed in Fig.~3(a).

Some remarks should be made on the obvious weaknesses of this simplistic model. First, it may be an oversimplification to assume a constant depinning current density. Especially near the Electron Beam Lithography- or Focussed Ion Beam-fabricated edges of the leads, local pinning centers should be expected. Second, no spontaneous redistribution of vortices is considered, as discussed in Supplemental Note~8. Finally, mutually and self-induced fields as well as edge effects are not taken into account.
As a consequence, these results are only considered as an indication that the critical state model provides the correct order of magnitude to explain the observed inverted hysteresis.

\newpage

\numsection{\\Inverted hysteresis and minor loops in a perpendicular magnetic field}\label{SI:sec:InvHystGaAs}

\begin{figure*}[tbh]
\centering
\includegraphics[width=\linewidth]{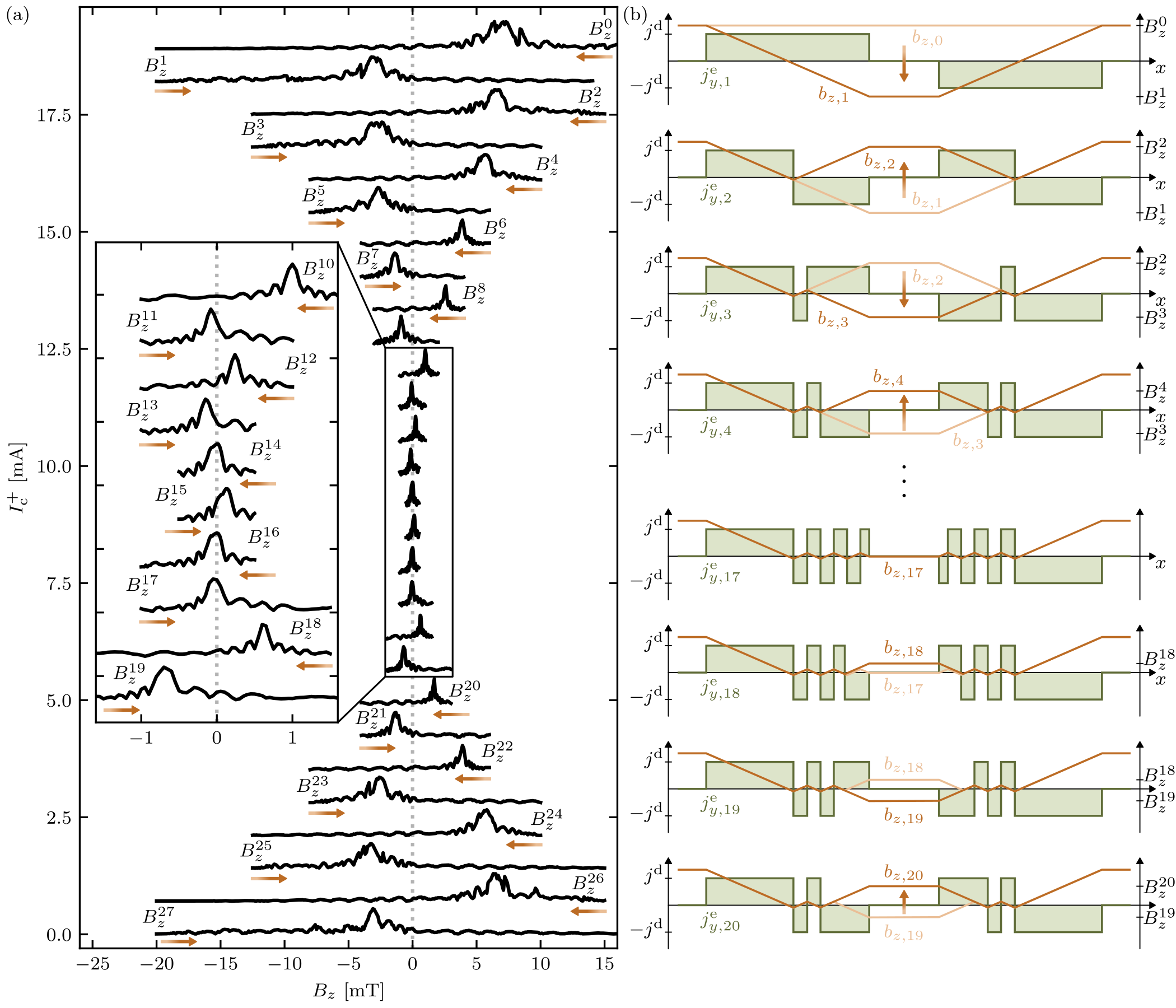}
\caption{Out-of-plane field minor loops in sample C. (a) $I_\textrm{c}(B_z)$ at different field ranges without field-cooling between sweeps (off-set for clarity). The sweep's starting point $B_z^i$ and direction are indicated for each trace. (b) Local magnetic field distribution $b_z(x)$ and edge current density $j_y^\textrm{e}(x)$ across the leads and the junction according to the critical state model outlined in the main text. The field distributions before ($b_{z,i-1}$, beige) and after ($b_{z,i}$, brown) an individual sweep are sketched together with the edge current density $j_{y,i}^\textrm{e}$ after the sweep, corresponding to $b_{z,i}$. For simplicity, the contribution of $b^\textrm{s}_z$ is omitted.}
\label{SI:fig:MinorLoops}
\end{figure*}

Further indication for the critical state model is found when gradually reducing the sweep range of the applied field. Minor loops of different ranges in $B_z$ are measured on sample C, starting from $\pm \SI{20}{mT}$ ($B_z^0$ and $B_z^1$) and decreasing towards $\pm \SI{0.5}{mT}$ ($B_z^{14}$ and $B_z^{15}$), see Fig.~\ref{SI:fig:MinorLoops}(a). No field-cooling is performed at the turn-around fields $B_z^i$. With reduced sweep range, the central lobes of the interference pattern shift closer together until the hysteretic effect vanishes and the central peak settles near zero field, as indicated by the gray dotted line. Upon increasing the field range from this state ($B_z^{16}$ onward), the lobe starts to shift away from zero and inverted hysteresis is recovered.

Fig.~\ref{SI:fig:MinorLoops}(b) describes a possible scenario for the local field distribution $b_z(x)$ under the premises of the critical state model. The following description is based on the lead to the left of the junction. The situation on the right-hand-side lead is antisymmetric, and the resulting stray fields of both sides add up on the area of the gap. After the initial field-cooling at $B_z^0$, the superconductor is in the fully penetrated vortex state and the field density $b_{z,0}$ is imprinted across the lead. As the field is swept towards the negative $B_z^1$, vortices leave the lead near the edges, creating a positive edge current distribution $j^\textrm{e}_{y,1}$, as sketched in the first graph of Fig.~\ref{SI:fig:MinorLoops}(b). As explained in the main text, this leads to inverted hysteresis effects. However, when returning to a positive field $B_z^2 < B_z^0$ (second graph in Fig.~\ref{SI:fig:MinorLoops}(b)) without performing a field-cooling routine to delete the vortex trapping history, the edge current is inverted only in the region of the lead that is closer to the junction. As a result, areas with positive and negative $j^\textrm{e}_y$ form. Since the region of negative $j^\textrm{e}_y$ is closer to the junction, the stray field resulting from there is larger at the junction area and can, therefore, still create inverted hysteresis, albeit smaller than in the initial case. Gradually decreasing the field sweep range further effectively folds $b_z(x)$, which translates to a spatially alternating $j^\textrm{e}_y(x)$. The resulting stray fields of each sector average out, leaving no field component to counteract the external field and create an inverted hysteresis. As the field range is increased again, the innermost folds are straightened out by the increasing applied field and allow a re-building of the inverted hysteresis.

It is noted that for smaller field ranges, $I_\textrm{c}(B_z)$ changes from a distorted and irregular shape to a more Fraunhofer-like pattern as the spontaneous jumps in $I_\textrm{c}$ disappear. The jumps return when the field range is increased again. This observation is consistent with the above explanation of the jumps in $I_\textrm{c}$ since for small field sweep ranges, the applied field is not sufficient to disturb the vortex distribution by spontaneous depinning of vortices.

\newpage

\numsection{\\Alignment method for the devices}\label{SI:sec:alignment}

\begin{figure}[tbh]
\centering
\includegraphics[width=5in]{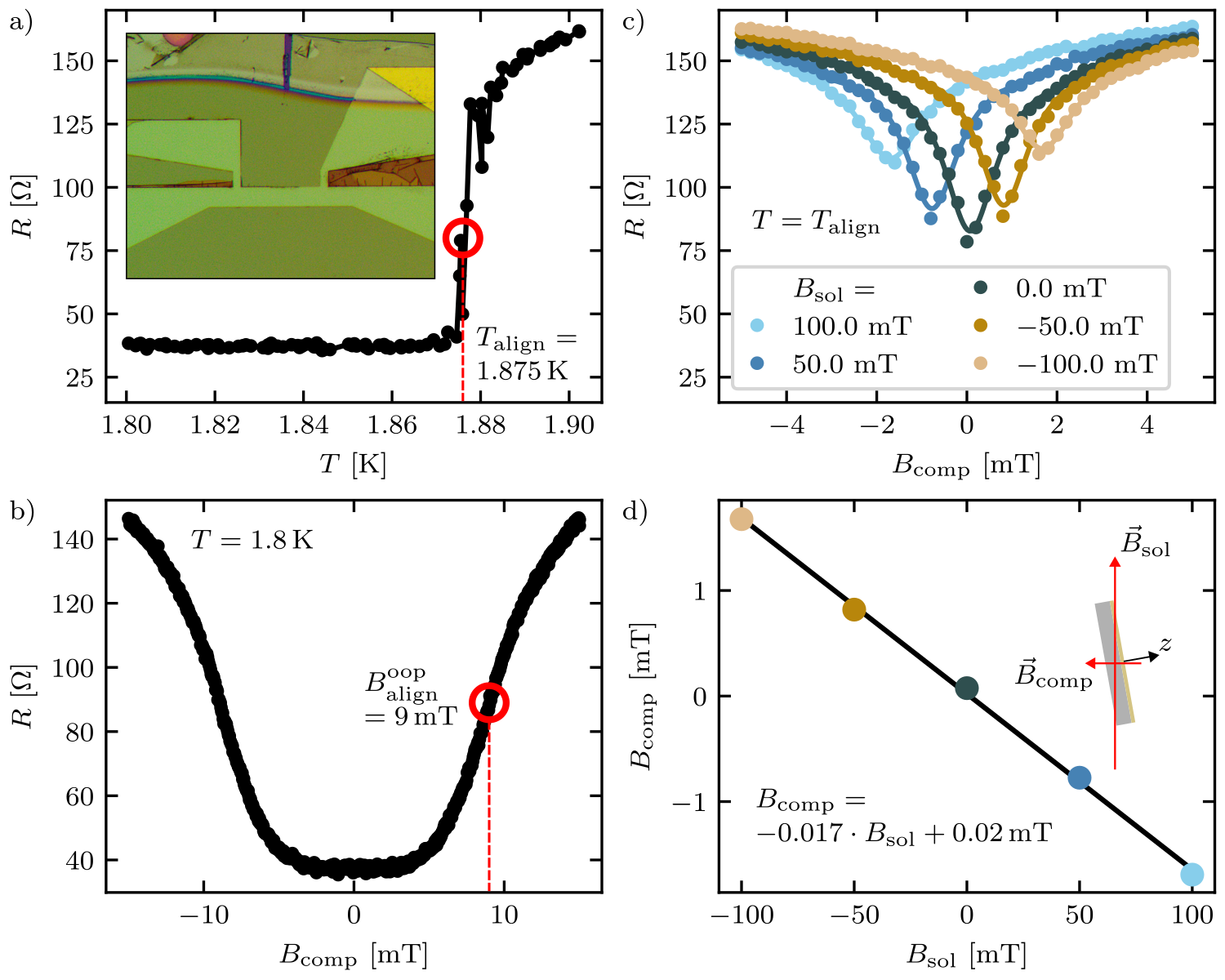}
\caption{Alignment procedure. (a) $R(T)$ of the Al stripe. The red circle highlights the temperature $T_\textrm{align}$ at which the in-plane alignment is performed. The inset shows a micrograph of the Al stripe in a four-point geometry. (b) Al stripe resistance under variable compensation field at $T=\SI{1.8}{K}$. The measurement is taken before the out-of-plane procedure with the sample aligned roughly perpendicular to $B_\textrm{comp}$. The red circle indicates the field at which out-of-plane alignment is conducted. (c) Typical in-plane alignment measurement conducted at $T=T_\textrm{align}$ for various solenoid fields. Solid lines indicate phenomenological Lorentzian fits to determine $B_\textrm{comp}$ at minimum stripe resistance. (d) Minimum positions extracted from fits in panel~(c). Linear fit yields parameters for alignment function $B_\textrm{comp} = a\cdot B_\textrm{sol} + b$. Inset shows orientation of sample in solenoid and compensation field.}
\label{SI:fig:alignment}
\end{figure}

As demonstrated by Fig.~3(b) in the main text, a small out-of-plane component caused by a sample misalignment of less than one degree from the field axis can lead to considerable narrowing of the central lobe in $I_\textrm{c}(B_\textrm{ip})$ and the formation of side lobes. These out-of-plane features mask the in-plane signature and may lead to a distortion in the diode factor. In order to optimize the in-plane alignment, the cryostat is equipped with a superconducting solenoid magnet used to apply the major component of the magnetic field $\vec{B}_\textrm{sol}$, as well as a pair of coils that generate an independent, perpendicular compensation field $\vec{B}_\textrm{comp}$. As a sensitive out-of-plane field probe, a \SI{10}{nm} thick Al stripe is evaporated on the substrate after the junction fabrication and patterned into a $\SI{200}{\micro m} \times \SI{50}{\micro m}$ stripe (see Fig.~\ref{SI:fig:alignment}(a) inset), located less than \SI{1}{mm} away from the junction device on the same chip. The longitudinal resistance of the stripe is measured using lock-in detection in a four-terminal geometry. Since the Al is a thin film type-I superconductor, its state is dominantly determined by the out-of-plane magnetic field and free of vortex-driven  effects.

Initially, the temperature-dependence of the stripe resistance is determined at zero applied field, as shown in Fig.~\ref{SI:fig:alignment}(a). The field compensation protocol operates at a temperature just below $T_\textrm{c}$ where the Al stripe is most sensitive to any applied magnetic field. From Fig.~\ref{SI:fig:alignment}(a), this alignment temperature is determined as $T_\textrm{align}=\SI{1.875}{K}$. To align the sample plane normal to the compensation field, the stripe resistance is measured as a function of $B_\textrm{comp}$ at a temperature well below $T_\textrm{c}$, see Fig.~\ref{SI:fig:alignment}(b), with the sample being roughly aligned with the nominal field axis. The field is set to a constant value of around \SI{9}{mT} determined by the point of the largest slope in resistance. The sample is then aligned perpendicular to $\vec{B}_\textrm{comp}$ by rotating the insert from outside the cryostat and maximizing the stripe resistance, corresponding to the largest out-of-plane component produced by $B_\textrm{comp}=\SI{9}{mT}$. Since this is an iterative trial-and-error process, no data is available on this step. The angular accuracy of this method is estimated to be around $5^\circ$. These steps are performed once in the beginning of each cool-down.

To access various in-plane angles $\theta$, the sample is mounted on a piezo rotator. Since the axis of the rotator is not guaranteed to be perpendicular to the sample plane, the following steps of the in-plane alignment have to be performed at each angle $\theta$. An exemplary alignment measurement is shown in Fig.~\ref{SI:fig:alignment}(c). At $T_\textrm{comp}$, $R(B_\textrm{comp})$ is measured for various constant values of the solenoid field $B_\textrm{sol}$. The minimum resistance of each curve represents the compensation field $B_\textrm{comp}$ necessary to counteract the spurious out-of-plane component of the applied $B_\textrm{sol}$. A Lorentzian fit function is used to identify the minimum position. (It should be noted that there is no theoretical model supporting the use of this function; the Lorentz fit is merely used to interpolate the true minimum position between a limited number of data points.) The extracted compensation fields are plotted in Fig.~\ref{SI:fig:alignment}(d) and fitted linearly, providing a function $B_\textrm{comp} = a\cdot B_\textrm{sol} + b$ to determine $B_\textrm{comp}$ needed to compensate the out-of-plane component of a given $B_\textrm{sol}$ during the field-dependent IV curves of the Josephson device under investigation.

The inset of Fig.~\ref{SI:fig:alignment}(d) schematically depicts the relative orientation of solenoid field $\vec{B}_\textrm{sol}$, compensation field $\vec{B}_\textrm{comp}$ and sample normal $z$. Note that the notation $B_\textrm{ip}$ and $B_z$ is relative to the sample plane.

Instead of using the compensation function to align the applied field in the sample plane, one can also extract the misalignment angle $\epsilon$ between the sample and $\vec{B}_\textrm{sol}$ by $\epsilon = \arctan(a)$. The data shown in Fig.~3(b) of the main text is acquired by performing the alignment procedure to calculate the misalignment angle $\epsilon$, but applying only $B_\textrm{sol}$ without compensation.

\newpage

\numsection{\\Inverted hysteresis in sample A in a perpendicular magnetic field}\label{SI:sec:InvHystGaAs}

\begin{figure}[tbh]
\centering
\includegraphics[width=\linewidth]{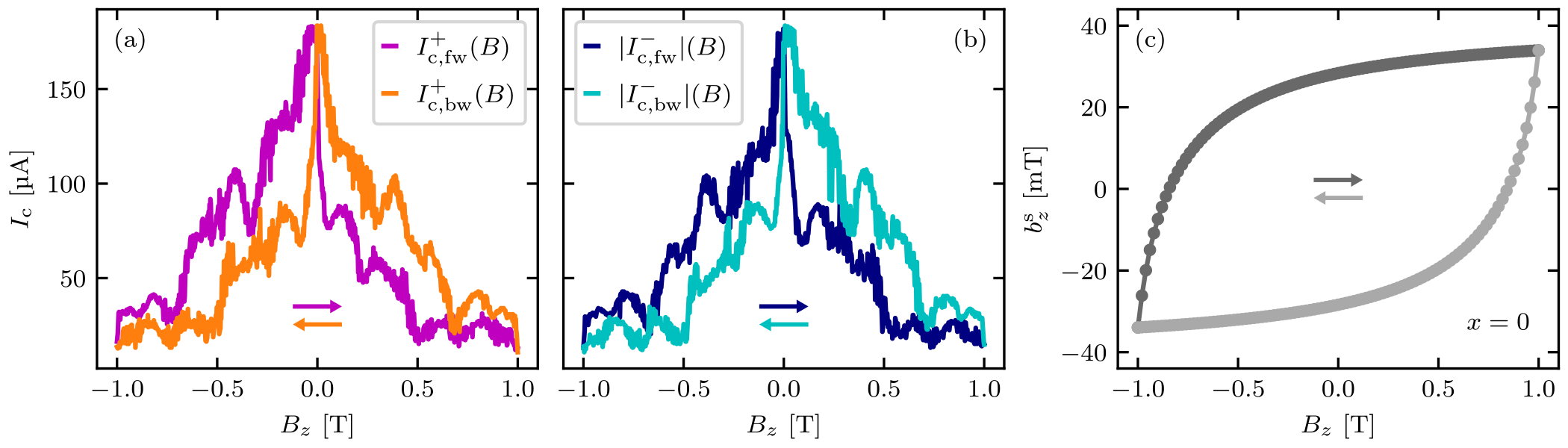}
\caption{Inverted hysteresis of $I_\textrm{c}$ curves for sample A under $B_z$. (a) Comparison of $I_\textrm{c}^+$ between opposite sweep directions corresponding to the main text Fig.~2. The colored arrows indicate the respective sweep directions. (b) Comparison of $I_\textrm{c}^-$ between opposite sweep directions, analogously to (a). (c) Additional field component $b_z^\textrm{s}$ simulated in the center of the junction ($x=0$) for both sweep directions without field-cooling.}
\label{SI:fig:InvHystGaAs}
\end{figure}

Figs.~\ref{SI:fig:InvHystGaAs}(a) and (b) compare $I_\textrm{c}^+$ and $I_\textrm{c}^-$ of sample A between the sweep directions of $B_z$, respectively, showing an apparent inverted hysteresis. The horizontal separation of side features is on the order of \SI{200}{mT}. For the estimation of the additional flux from the stray field $b_z^\textrm{s}$ resulting from the critical state model, the approach outlined in Supplemental Note~9 is adapted to the case without field-cooling by adding a negative contribution from the current distribution built up during the previous sweep. Following Eq.~(\ref{SI:eq:xmax}), the horizontal position of the sign change depends on the applied field $B_z$ as $x^\textrm{sng} = |B_z^0 - B_z| \, / \, 2\mu_0 j^\textrm{d}$. Using $B_z^0=\pm\SI{1}{T}$, $B_z \in [-1,+1]$ T and the same critical depinning current density $j^\textrm{d}$ for the Nb leads as for sample B, one obtains $x^\textrm{max} = \SI{1.6}{\micro m}$. Since this is much larger than the junction length of $\sim \SI{20}{nm}$, $b_z^\textrm{s}$ is homogeneous across the length of the junction. Hence, it is sufficient to evaluate Eq.~(\ref{SI:eq:NumInt}) only in the center of the junction at $x=0$. The result is shown in Fig.~\ref{SI:fig:InvHystGaAs}(c) as a function of $B_z$ for both sweep directions. This suggests that, for the chosen parameters, the critical state model can account for an inverted hysteresis of the order of \SI{50}{mT}. Taking into account the simplifications discussed in Supplemental Note~9, the quantitative deviation by a factor of $\sim 4$ is considered small enough to accept the qualitative explanation of the observed inverted hysteresis by the critical state model.

With respect to the Fe-layer of the stack~\cite{Bayreuther_2012, Chen_2016, Chen_2024}, it should be noted that experiments on magnetic Josephson devices have demonstrated hysteretic Fraunhofer patterns caused by the stray field of the remnant vertical magnetization keeping the flux through the junction area finite even when the applied field has reached zero~\cite{Larkin_2012, Bolginov_2012, Baek_2014}. In the presented case, the hysteresis appears to be inverted and therefore is unlikely to be related to magnetization switching of the Fe layer.

\newpage

\bibliography{00_arXiv_references.bib}